\documentclass{aastex62}
\usepackage{graphicx}
\graphicspath{{figures/}}
\usepackage{amsmath}
\usepackage{amssymb}
\usepackage{color}
\submitjournal{AJ}
\accepted{May 14, 2019}
\shorttitle{Io's Volcanic Activity: 2013-2018}
\shortauthors{de Kleer et al.}
\begin{document}

\title{Io's Volcanic Activity from Time Domain Adaptive Optics Observations: 2013-2018}

\correspondingauthor{Katherine de Kleer}
\email{dekleer@caltech.edu}

\author[0000-0000-0000-0000]{Katherine de Kleer}
\affil{California Institute of Technology \\
1200 E California Blvd M/C 150-21 \\
Pasadena, CA 91125, USA}
\author{Imke de Pater}
\affil{University of California, Berkeley, Berkeley CA}
\author{Edward M. Molter}
\affil{University of California, Berkeley, Berkeley CA}
\author{Elizabeth Banks}
\affil{The Pembroke Hill High School, Kansas City MO}
\author{Ashley Gerard Davies}
\affil{Jet Propulsion Laboratory, California Institute of Technology, Pasadena CA}
\author{Carlos Alvarez}
\affil{W. M. Keck Observatory, Waimea HI}
\author{Randy Campbell}
\affil{W. M. Keck Observatory, Waimea HI}
\author{Joel Aycock}
\affil{W. M. Keck Observatory, Waimea HI}
\author{John Pelletier}
\affil{W. M. Keck Observatory, Waimea HI}
\author{Terry Stickel}
\affil{W. M. Keck Observatory, Waimea HI}
\author{Glenn G. Kacprzak}
\affil{Swinburne University of Technology, Hawthorn, Victoria, Australia}
\author{Nikole M. Nielsen}
\affil{Swinburne University of Technology, Hawthorn, Victoria, Australia}
\author{Daniel Stern}
\affil{Jet Propulsion Laboratory, California Institute of Technology, Pasadena CA}
\author{Joshua Tollefson}
\affil{University of California, Berkeley, Berkeley CA}

\begin{abstract}
We present measurements of the near-infrared brightness of Io's hot spots derived from 2-5 $\mu$m imaging with adaptive optics on the Keck and Gemini N telescopes. The data were obtained on 271 nights between August 2013 and the end of 2018, and include nearly 1000 detections of over 75 unique hot spots. The 100 observations obtained between 2013 and 2015 have been previously published in de Kleer and de Pater (2016a); the observations since the start of 2016 are presented here for the first time, and the analysis is updated to include the full five-year dataset. These data provide insight into the global properties of Io's volcanism. Several new hot spots and bright eruptions have been detected, and the preference for bright eruptions to occur on Io's trailing hemisphere noted in the 2013-2015 data (de Kleer and de Pater 2016a) is strengthened by the larger dataset and remains unexplained. The program overlapped in time with \textit{Sprint-A/EXCEED} and \textit{Juno} observations of the jovian system, and correlations with transient phenomena seen in other components of the system have the potential to inform our understanding of the impact of Io's volcanism on Jupiter and its neutral/plasma environment. 
\end{abstract}


\section{Introduction}
Io's dramatic volcanic activity exhibits a high degree of spatial and temporal variability. The distribution of volcanic thermal emission in space and time contains information on the underlying volcanic advection processes, providing a window into the nature of Io's geological processes as well as into how tidal heating impacts the characteristics of the volcanism it powers. \par
While some of Io's volcanoes have remained persistently active since the \textit{Voyager} fly-bys in 1979, numerous transient eruptions appear and subside in a matter of days, hours, or even minutes (e.g. Johnson et al. 1988; Veeder et al. 1994; de Pater et al. 2014; de Kleer et al. 2014; Tsang et al. 2014; Davies et al. 2018). The timeline of thermal activity for a given volcano is indicative of the style of volcanism and hence geological processes active at that site (Davies et al. 2010). The time intervals between eruptions at a given site can provide information on characteristic resupply timescales, while a comparison of eruption timing between sites has the potential to illuminate eruption clustering if present. Finally, the periodic forcing of Io may translate into specific temporal signatures that may be apparent in thermal timelines. \par
Volcanoes are also distributed non-randomly across Io's surface, showing in particular a dearth of activity at the sub- and anti-jovian longitudes, as well as in polar regions (Hamilton et al. 2013; Veeder et al. 2015; de Kleer and de Pater 2016b), although no dataset published to date has had good coverage of the high latitudes. The spatial distribution of Io's surface heat flow may place constraints on models for tidal heat dissipation in Io's interior, or may indicate the degree of fluid flow in Io's mantle through the amount of smoothing in the observed spatial trends relative to the expected patterns. Without allowing for lateral movement of melt, the end-member case of heat deposition in a shallow aesthenosphere predicts higher heat flow at lower latitudes with the greatest heat flow centered at the sub-jovian and anti-jovian regions. In contrast, the  end-member case of deep mantle heating results in enhanced heat flow at the poles (Gaskell et al., 1988; Segatz et al., 1988). \par
Determining the temporal and spatial distribution of Io's volcanism requires a large sample size of hot spot detections over a range of timescales. We have been building up a database of thermal emission from individual volcanoes on Io's surface since 2013, when we initiated a time domain campaign of adaptive optics imaging of Io's volcanoes at the Keck and Gemini N telescopes. Io has been observed using adaptive optics on Keck since 2001 (Marchis et al. 2002), but only since 2013 has there been a dedicated Io observing program at such high cadence. These observations spatially resolve Io, permitting the identification of individual active volcanoes, and are often made at multiple wavelengths in the 2-5 $\mu$m range in order to constrain temperature and total power output. The observations have a typical spatial resolution of $\sim$100-500 km depending on telescope, wavelength, and sky conditions. The collective dataset is well suited to an investigation into the volcanic eruption processes at individual hot spots, which requires data capturing the time evolution of the eruptions, and to identification of spatial and temporal patterns in the distribution of activity. \par
The prior Io dataset that is most comparable in cadence, wavelength, and spatial resolution is from the \textit{Galileo} Near-Infrared Mapping Spectrometer (NIMS; Carlson et al. 1992), which observed Io on 25 distinct passes with a typical spatial resolution of 100-400 km on Io's surface, including some images with resolutions as coarse as 725 km and as fine as 100 meters (see Table 3.2 in Davies 2007). NIMS detected thermal emission from 115 unique hot spots (Davies et al. 2012; Veeder et al. 2012; 2015), each detected between one and 50$+$ times over the course of the mission. Long-term programs observing Io's thermal emission from the NASA InfraRed Telescope Facility have also been very successful (Spencer et al. 1990; Veeder et al. 1994; Rathbun and Spencer 2010). While such data do not spatially resolve Io, techniques such as lucky imaging and observing Io as the satellite enters or emerges from occultation behind Jupiter have permitted brightness measurements of individual volcanoes. Though sensitive only to the brightest events and only to the Jupiter-facing hemisphere, occultation observations have by far the longest time baseline, having been made on more than 100 occasions over the past $>$2 decades (Rathbun et al. 2018), albeit at only one wavelength (3.5 or 3.8 $\mu$m). \par
Our spatial resolution and sensitivity to faint hot spots is intermediate between NIMS data and occultation observations, and is comparable to a typical NIMS observation. Our cadence and total number of observations are higher than all prior datasets, although the time baseline of our high cadence campaign is much shorter than the decadal timescales covered by the occultation datasets. \par
Our campaign is introduced in de Kleer et al. (2014), and the analysis methods and results from the first 2.5 years of the program (100 nights of observation) are given in de Kleer and de Pater (2016a). Here we present results from the 2016-2018 observations, and a joint analysis of all data to date, 2013-2018. The flexible scheduling capabilities at Gemini N, and our Twilight Zone observing program at Keck\footnote{https://www2.keck.hawaii.edu/inst/tda/TwilightZone.html}, have been instrumental in achieving the high cadence and quantity of observations.  The observations and data analysis methods are reviewed in Section \ref{sec:obs}, the results are presented and discussed in Section \ref{sec:results}, and the conclusions are summarized in Section \ref{sec:conc}.
\section{Observations and data analysis} \label{sec:obs}
\begin{figure}
\centering
\includegraphics[width=18cm]{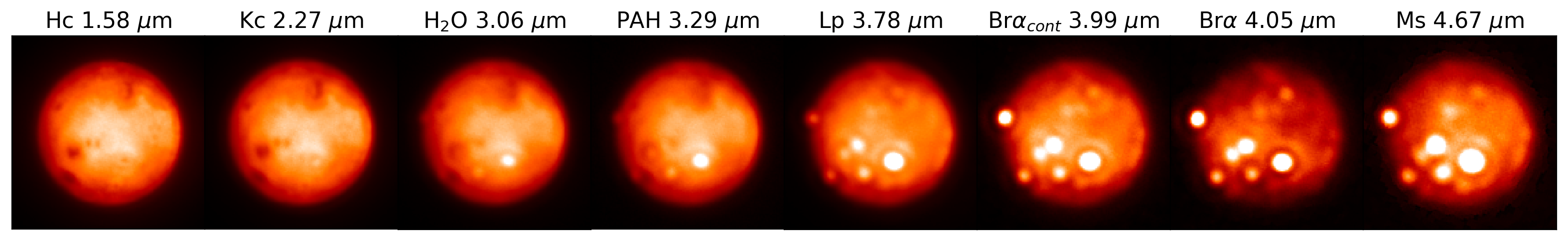}
\caption{Images from Keck on 2017 May 28 demonstrating the range of filters used in the observations. All images were taken within a 30-minute window, and each is labeled with the filter name and central wavelength. \label{fig:keckims}}
\end{figure}

We observed Io in the near-infrared with adaptive optics on 271 nights between August 2013 and July 2018; the observing dates and details are given in Table \ref{tbl:obs}. Observations were made with the NIRI imager on Gemini N (Hodapp et al. 2003) combined with the ALTAIR adaptive optics system in Natural Guide Star (NGS) mode, and with the NIRC2 imager on Keck II also using NGS adaptive optics (Wizinowich et al. 2000). The Gemini N data constitute 80\% of the total visits, and include images in the L' (3.78 $\mu$m) and K-cont (2.27 $\mu$m) filters. The Keck images were taken in a variety of filters from H-cont (1.58 $\mu$m) to Ms (4.67 $\mu$m), shown in Figure \ref{fig:keckims}. \par
Images are flux calibrated to a standard star if a star was observed and the night was photometric; otherwise the images are calibrated to volcano-free regions of Io's disk, which do not change measurably with time. Within each image, all hot spots are identified; their pixel locations are translated to latitude and longitude coordinates on Io's disk based on Io's ephemeris; and their intensity is measured based on an aperture photometry approach adapted to point sources on a bright background (de Pater et al. 2014). All observing, data reduction, and analysis procedures are described in detail in de Kleer and de Pater (2016a), and an identical approach is used here. Of the full dataset, the 100 observations from 2013 through the end of 2015 were published in de Kleer and de Pater (2016a), while the 171 observations from 2016-2018 are presented here for the first time. \par
The detection limits for hot spots in the Keck and Gemini N images are given as a function of emission angle in Appendix A of de Kleer and de Pater (2016a). We use these limits to define the sensitivity of the dataset as a whole to hot spots at different longitudes. This sensitivity varies by up to 20\% across Io's surface, and is used to correct the longitudinal hot spot distribution described in Section \ref{sec:spatdist}.
\section{Results \& Discussion} \label{sec:results}
\begin{figure}
\centering
\includegraphics[width=16cm]{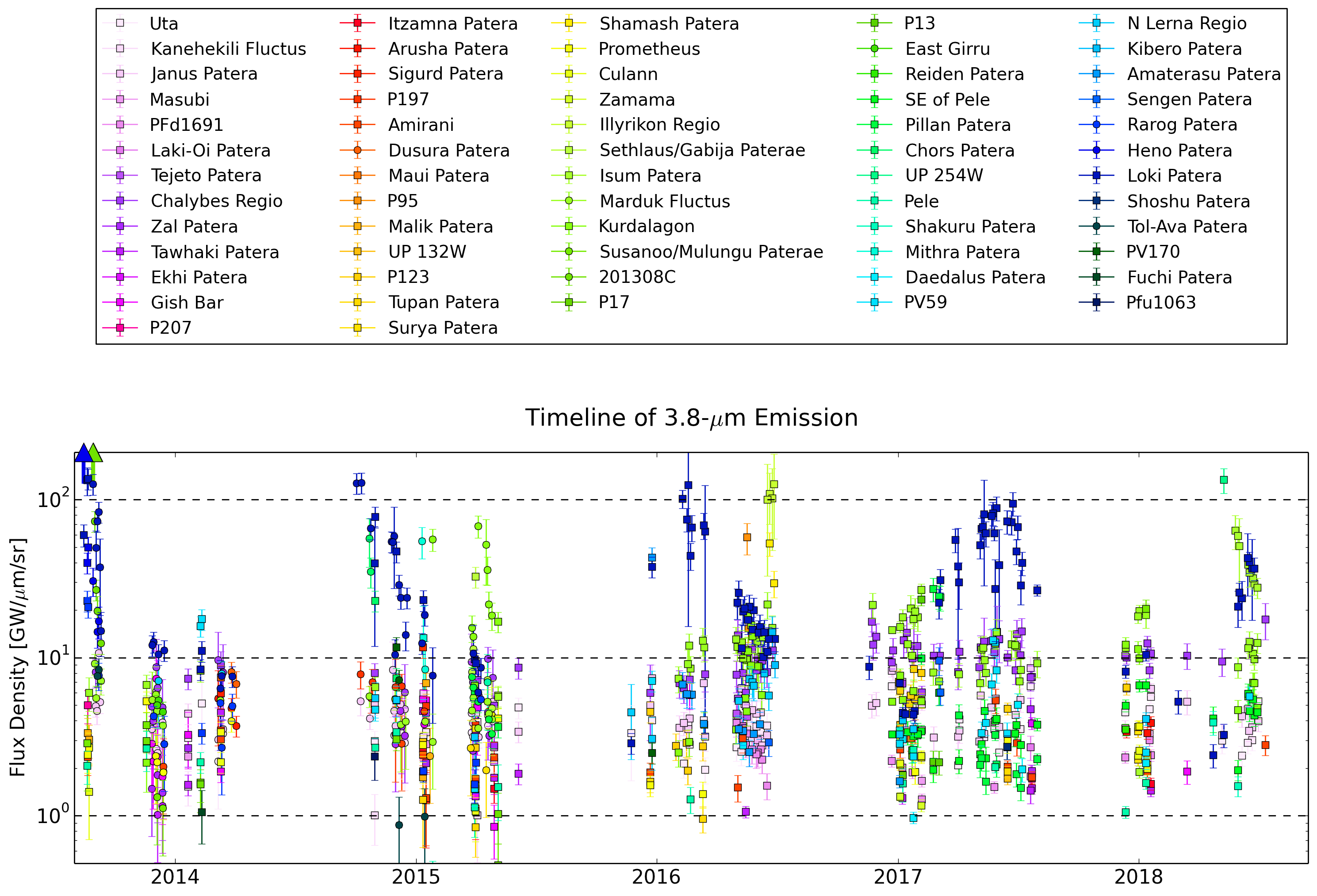}
\caption{Timeline of Io's volcanic activity from 2013-2018. All L'-band detections of thermal emission from volcanic centers are plotted, with each volcanic center in a different color. The gaps in the timeline correspond to periods when the Jupiter system was not observable from Maunakea. The timeline shows that there are multi-month intervals with no bright activity, and other intervals when several large eruptions took place. \label{fig:timeline}}
\end{figure}

The coordinates, number of detections, and average brightness of each of the 75 hot spots detected and tracked by our program are listed in Table \ref{tbl:overview}. The full set of near-infrared brightnesses in all filters for all 980 hot spot detections are tabulated in Table \ref{tbl:hsphot}. In some cases, the location of a hot spot appears to shift over time or transition from one active site to another nearby; in cases where it is not clear from the data whether the emission over time is produced by a single site or multiple nearby sites, we tabulate all detections under a single site name. The timeline of L'-band (3.78 $\mu$m) brightness of all volcanoes is shown in Figure \ref{fig:timeline}, which gives a sense for the global variability of Io's volcanism over this period.
\subsection{Energetic eruptions}
Since 2013 we have detected bright eruptions at eighteen sites, where we define ``bright'' as a maximum L'-band brightness greater than 20 GW/$\mu$m/sr. There is no hot spot on Io that consistently exhibits this level of activity, and this cut-off therefore selects for transient events. These eruptions are typically vigorous, high-power events with significant short-wavelength emission. The majority of these were short-lived, exhibiting their peak brightness for only a few days before decaying. A few volcanoes are exceptions to this rule, producing thermal emission that is consistently present at a moderate level while also exhibiting infrequent brightenings; these volcanoes are Loki Patera, Pillan Patera, Marduk Fluctus, and Kurdalagon Patera. The first three of these were active throughout the period of observation, while Kurdalagon Patera was not detected before its eruption at the beginning of 2015 but was subsequently active and variable through 2018. \par
Table \ref{tbl:transients} lists these eighteen hot spots and the brightest L'-band intensity measured at each during our program. Note that only the single brightest detection is given even though some volcanoes had multiple large eruptions. The full timeline for each of these hot spots can be found in Table \ref{tbl:hsphot}. The events that occurred prior to the end of 2015 were presented in de Kleer and de Pater (2016a); our discussion here therefore focuses on eruptions detected since the beginning of 2016. \par
For volcanoes with detections at multiple wavelengths on a given night, we fit a Planck spectrum to estimate the temperature. All detections with measured temperatures above 800 K are given in Table \ref{tbl:highT}; in total there are 32 such detections at 18 unique hot spots. These hot spots are not exactly the same set as the 18 sites where bright eruptions are seen, although there is significant overlap. While this is a small fraction of the total number of observations for which we were able to derive temperature estimates, it confirms previous findings that these high temperatures are common and widespread across a variety of volcanic styles and are not limited to outburst events (e.g. Carr 1986; Lopes-Gautier et al. 1999). However, we note that Io's active volcanoes likely exhibit a range of temperatures from near the magma temperature ($\sim$1500 K if the magma is basaltic) down to near the passive surface temperature ($\sim$125 K), and the temperatures recovered in a single temperature fit therefore do not directly represent any physical temperature (although they do serve as a lower limit on the eruption temperature). In fact, if the magma composition is the same at all of Io's volcanoes, then the best-fit temperature instead reflects the proportion of high-temperature to low-temperature emitting areas, and high fitted temperatures are indicative of volcanic eruptions vigorous enough that sufficient area is exposed at very high temperatures to yield a short wavelength peak in thermal emission (e.g. Davies et al. 2010). \par
The temperatures in Table \ref{tbl:highT} are derived as in de Kleer and de Pater (2016a), using Markov Chain Monte Carlo simulations to determine the probability distribution for temperature and emitting area, from which the uncertainties are also derived. Measurements from all available wavelengths are used, incorporating uncertainties on the intensity measurements, and a maximum K-cont (2.27 $\mu$m) brightness limit of 7 GW/$\mu$m/sr is imposed in the fitting when the hot spot was not detected at that wavelength. \par
The temperature estimates are derived from the intensity measurements given in Table \ref{tbl:hsphot}, which have been corrected for geometric foreshortening. However, in the case of a high emission angle observation of an event of significant vertical extent such as fire fountaining, the short-wavelength emission may arise primarily from the hot fountaining component that is not foreshortened, while the longer-wavelength emission arises from both the fountains and the resultant lava flows, which are foreshortened. Applying the foreshortening correction across all wavelengths may therefore inflate the derived short-wavelength emission and hence the temperature, so that temperatures derived from high emission angle observations should be viewed with caution. \par
\subsubsection{Eruption at P95 (May 2016)}
In May 2016 a bright and short-lived eruption was detected at patera P95, near 10$^{\circ}$S 128$^{\circ}$W. The eruption was first detected on May 17 with a temperature around 1000 K. The second and final detection of the eruption occurred two days later on May 19, and the eruption had already declined significantly in brightness by this time. The latest non-detection of the site prior to the eruption was May 12, while the eruption had faded to below $I_{Lp}\sim$5 GW/$\mu$m/sr by May 24, and to below the detection limit even at optimal viewing geometry ($I_{Lp} \sim$3 GW/$\mu$m/sr) by May 28. While high in both temperature and infrared emission, this event therefore was short-lived, detected only over a 3-night period and constrained to be active at a detectable level for less than 16 days. 
\subsubsection{Eruptions at Shamash Patera and in the Illyrikon Regio (June 2016)}
\begin{figure}
\centering
\includegraphics[width=12cm]{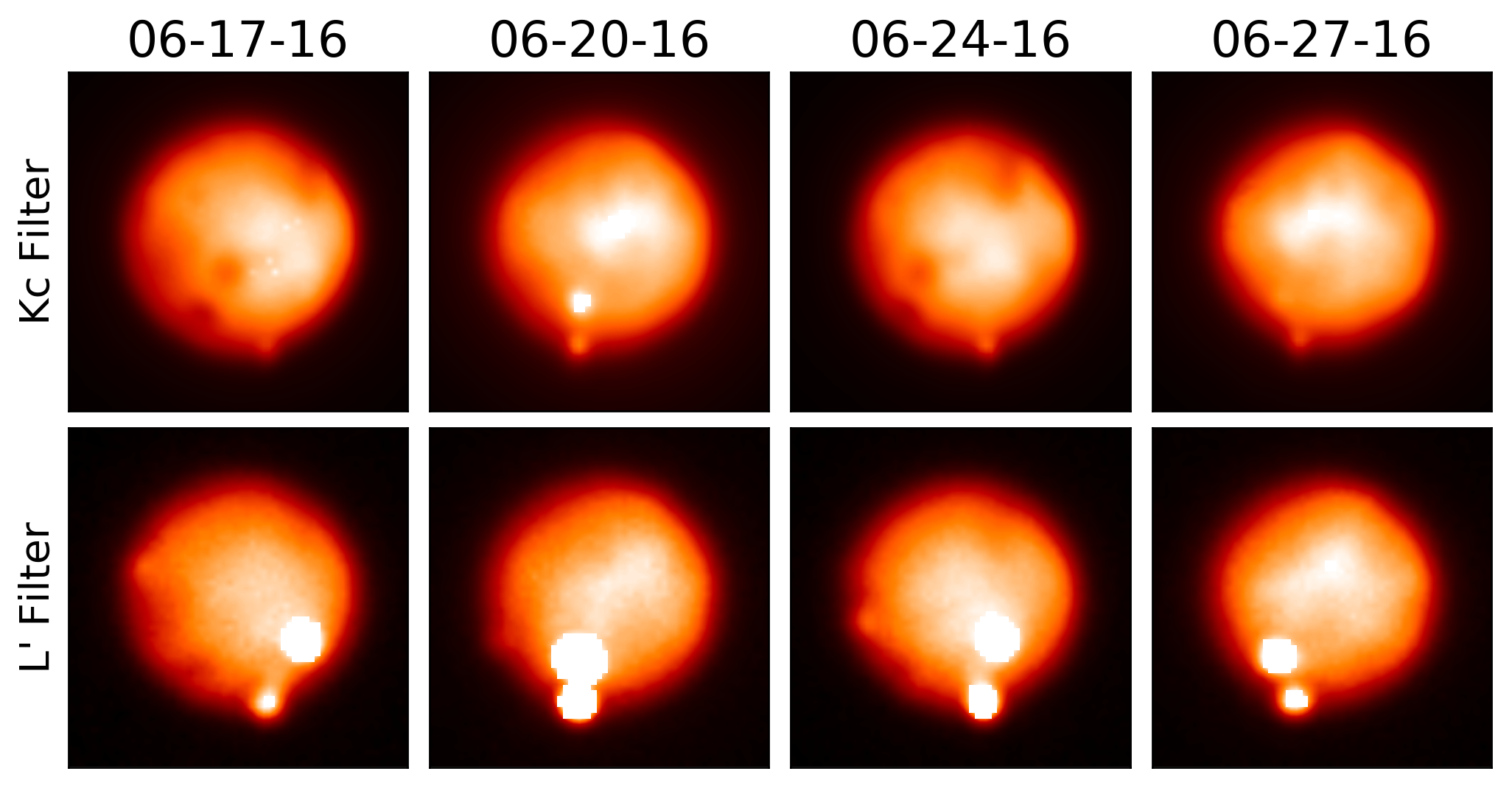}
\caption{Near-infrared images from Gemini N of eruptions in Io's southern hemisphere in 2016. The hot spot near the south pole is at a new site in the Illyrikon Regio. Despite the apparent similarity between all four L' images, the viewing geometry changes significantly between observations: from left to right the central meridian longitudes are 249$^{\circ}$; 138$^{\circ}$; 233$^{\circ}$; and 123$^{\circ}$ W. The mid-latitude hot spot is Marduk Fluctus on June 17 and 24, and is Shamash Patera on June 20 and 27.  \label{fig:illsham}}
\end{figure}

A pair of dramatic eruptions occurred in the southern hemisphere at Shamash Patera (33$^{\circ}$S 150$^{\circ}$W) and in Illyrikon Regio near 71$^{\circ}$S 180$^{\circ}$W in June 2016. The eruption in Illyrikon Regio was first detected on June 17, and began no earlier than June 10. Shamash Patera was still not active as late as June 18, after the eruption at Illyrikon Regio had begun, but exhibited bright activity on June 20. The eruption at Shamash Patera had decayed and cooled somewhat by June 27, while the eruption in Illyrikon Regio stayed bright and hot through the end of June, after which we had no further observations until November. Figure \ref{fig:illsham} shows images of the eruptions at these volcanoes. The two volcanoes appear close in the images but are separated by over 1000 km of surface distance. The location of the hot spot in Illyrikon Regio is poorly constrained due to the high emission angle of all observations and we cannot conclusively match a surface feature at its location, but the positioning of a dark patera at 71$^{\circ}$S 170$^{\circ}$W is consistent with some of the thermal emission detections, whose best-fit longitudes fall in the range of 165-193$^{\circ}$W. At 71$^{\circ}$S, this is the most polar hot spot detected by our program.
\subsubsection{Eruption at UP 254W (May 2018)}
On May 10, 2018 a bright, high-temperature ($\sim$1000 K) eruption was detected at 37$^{\circ}$S 254$^{\circ}$W; a small patera at exactly this location is seen in spacecraft surface imaging (Williams et al. 2011a) and is a plausible source of the eruption. The hot spot was detected again on May 31 but had dimmed nearly to invisibility, and was not seen again. Although the hot spot location is close to the hot spot we refer to as ``SE of Pele'', these hot spots are clearly distinct and are spatially resolved in the May 31 observations. No prior activity at this location has been documented.
\subsubsection{Eruption at Isum Patera (May-June 2018)}
\begin{figure}
\centering
\includegraphics[width=18cm]{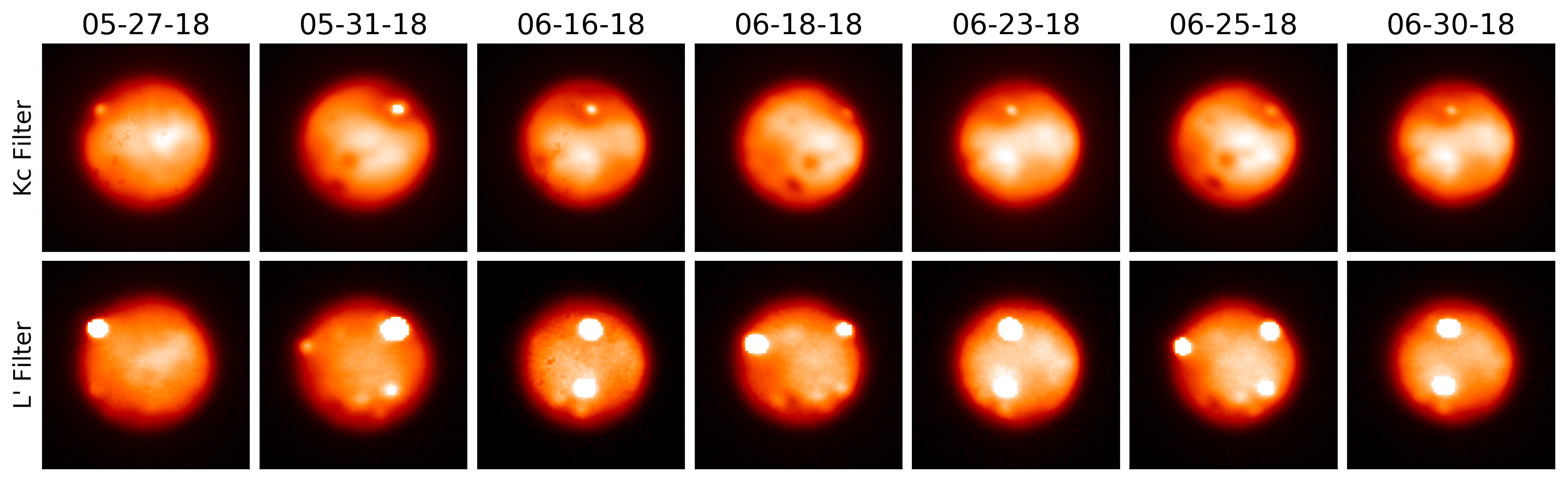}
\caption{Near-infrared images from Gemini N of the eruption at Isum Patera in spring 2018. The eruption is the only hot spot visible in the K filter (2.27 $\mu$m), and is seen at a corresponding location in the L' images (3.78 $\mu$m). The bright hot spot south of Isum Patera is Marduk Fluctus. \label{fig:isum}}
\end{figure}

\begin{figure}
\centering
\includegraphics[width=16cm]{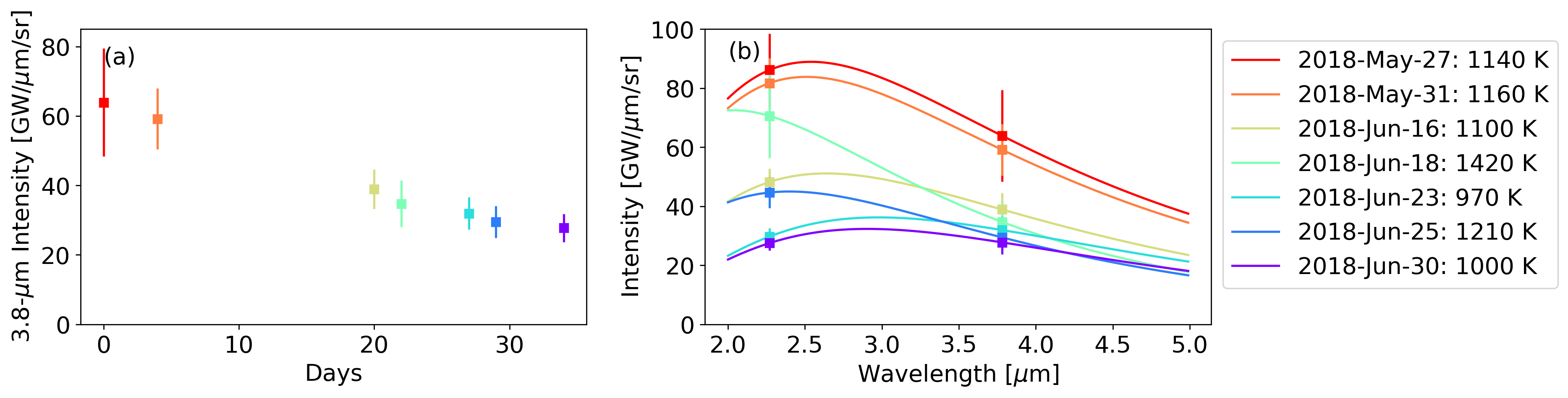}
\caption{Eruption at Isum Patera in 2018. (a) Timeline of 3.8 $\mu$m intensity over a $\sim$1-month period; (b) Temperature fits to the 2.3- and 3.8-$\mu$m measured brightnesses. Temperatures may be over-estimated if lava fountaining is occurring, as discussed in the text.\label{fig:isum2}}
\end{figure}

Of the high-power events detected in 2016-2018, the most dramatic in both temperature and duration was an eruption at Isum Patera in May-June 2018. The event began prior to May 27 and exhibited temperatures around or above 1000 K for the subsequent month. The total emission decayed steadily over this period, suggesting that new magma was being erupted throughout but at a rate that decreased with time. Figures \ref{fig:isum} and \ref{fig:isum2} show images of the eruption and plot its infrared timeline and the corresponding temperature fits.
\subsection{Activity at new hot spots}
Many of Io's most active hot spots today have exhibited persistent or episodic activity back to the \textit{Voyager} and \textit{Galileo} missions, nearly 40 years in some cases. However, the detection of new hot spots at locations where no thermal emission was previously seen is also common in the ground-based datasets (de Pater et al. 2016; Cantrall et al. 2018), including hot spots where no corresponding surface features are seen. The detection of these new hot spots improves our understanding of the distribution of active volcanic centers on Io's surface and of their heat flow. \par
Cantrall et al. (2018) identified 24 hot spots that had been detected in ground-based data and were not seen by \textit{Galileo}, more than a quarter of the total number of hot spots seen in the ground-based dataset. The new data presented here bring the total number of hot spots seen in the ground-based adaptive optics datasets from 2001-2018 to 104, 29 of which were not seen by previous spacecraft missions. In the new data presented here, the hot spots where thermal emission had not been previously detected were: Ekhi Patera; the hot spot in Illyrikon Regio; an unknown location near 54$^{\circ}$N 218$^{\circ}$W; the hot spot SE of Pele; and an unnamed patera near 37$^{\circ}$S 254$^{\circ}$W. Of these five, emission from Ekhi Patera and the unknown location at 218$^{\circ}$W was detected only once and at low brightness; the hot spot SE of Pele was consistently detected from Dec 2016 through the end of 2018 though at a low level; and the hot spots in Illyrikon Regio and at 254$^{\circ}$W were locations where bright eruptions took place. \par
Of the 75 hot spots detected in 2013-2018, about 1/3 of them were detected throughout the period of observation, 1/3 were detected only in 2013-2015, and 1/3 were detected only in 2016-2018. For the set of hot spots that were only detected during one of the two intervals of observation, the majority were detected less than half a dozen times. This characteristic, in combination with the fact that a substantial fraction of these hot spots were previously detected by \textit{Galileo} or \textit{Voyager}, suggests that despite the apparent turnover in activity between observing intervals, it is likely that nearly all hot spots have been geologically active throughout the period of space- and Earth-based observation, but that they only output surface thermal emission sporadically, so that the set of hot spots detected in an observation period may depend heavily on the exact timing of the observations. \par
A clear exception to this is the category of hot spots where no previous thermal emission has been seen, no clear patera feature is present at the site of the emission, and yet the hot spot stays persistently active for years after the activity is first seen. In this dataset, the two most prominent examples are the hot spot in Chalybes Regio, and the hot spot SE of Pele. These appear to be locations where volcanic activity initiated since the \textit{Galileo} mission. \par
Chalybes Regio is a northern region with extensive lava flow fields. Thermal emission was first detected from this location in 2010 (de Pater et al. 2014), and was attributed to PFu 2083, a small patera floor unit identified by Williams et al. (2011a) at 56$^{\circ}$N 74$^{\circ}$W. Thermal emission from this location was consistently detected in every observation we made between 2013 and 2018 that had appropriate viewing geometry. In many cases the emission appears spatially extended, indicative of multiple active areas that are not spatially resolved in the data. \par
The hot spot SE of Pele, located near 35$^{\circ}$S 240$^{\circ}$W, was first detected on Dec 23, 2016 and was thereafter detected through the end of our observation period in mid-2018. While there are many flows and paterae in this general area of Io's surface, there is no patera whose location provides a good match to the observed thermal emission.
\subsection{Persistent volcanoes and periodicities}
\begin{figure}
\centering
\includegraphics[width=16cm]{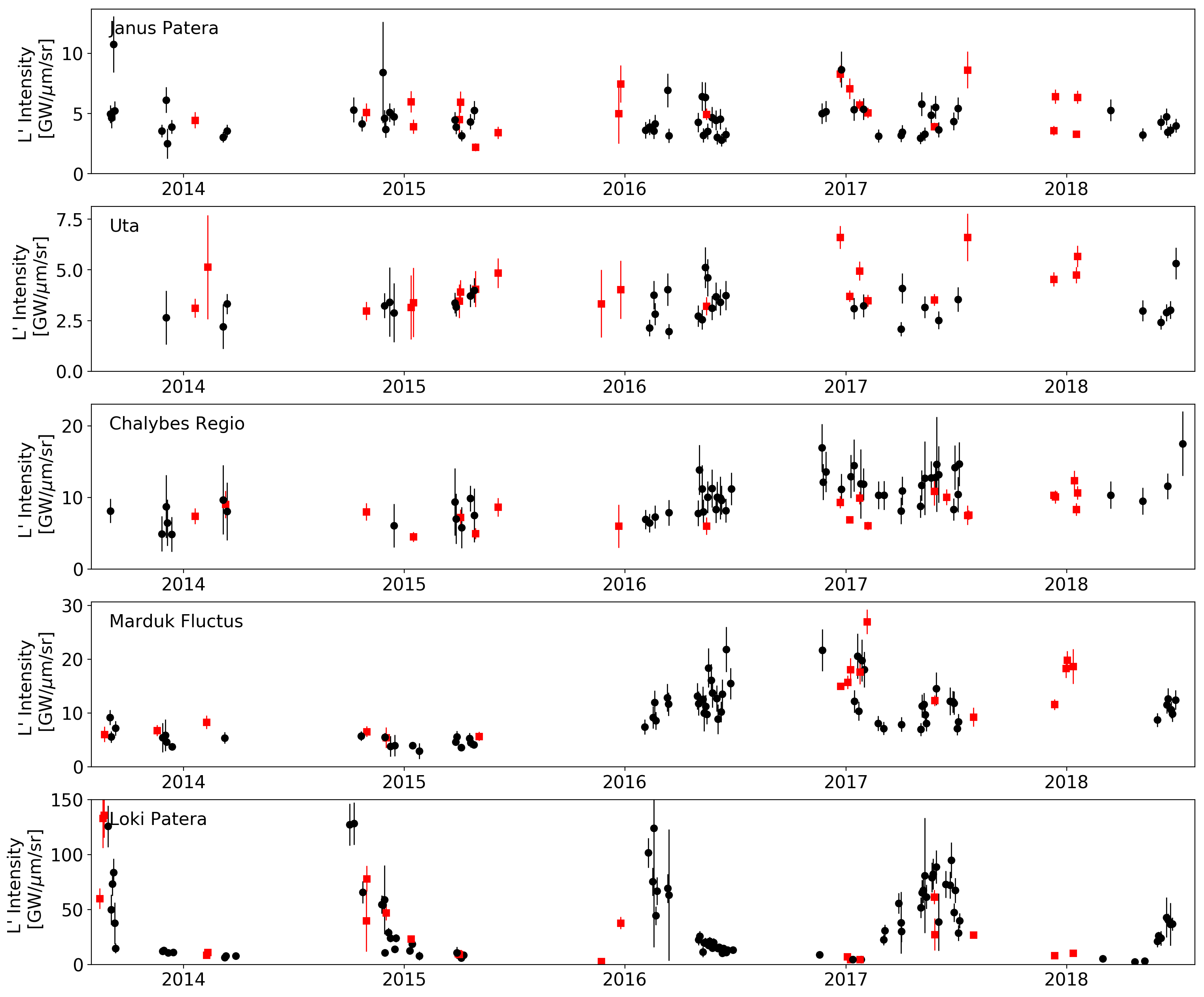}
\caption{Activity timelines for five persistently active hot spots.  Black circles and red squares indicate detections with Gemini N and Keck respectively. \label{fig:indivhs}}
\end{figure}

The sites that were most consistently detected during our campaign were Loki Patera (113 detections), Marduk Fluctus (87 detections), Janus Patera (84 detections), the hot spot in Chalybes Regio (80 detections), and Uta (57 detections). The first four of these were detected every time the viewing geometry was favorable. The hot spot at Uta does not appear to stay localized to the patera and may in fact be composed of multiple closely spaced volcanic centers not clearly resolved from one another. The timelines for these five hot spots are shown in Figure \ref{fig:indivhs}. The large number of observations of each of these hot spots provides a database of thermal brightness that may be used to fit models for volcanic activity style. In addition, the quantity and cadence of the images result in a dataset that is sensitive to periodicities in volcanic brightness on timescales from days to months. The tidal forcing that both powers the activity and deforms Io's crust is periodic, and the resultant activity may reflect these periodicities depending on the rheology and eruption mechanism. \par
\begin{figure}
\centering
\includegraphics[width=16cm]{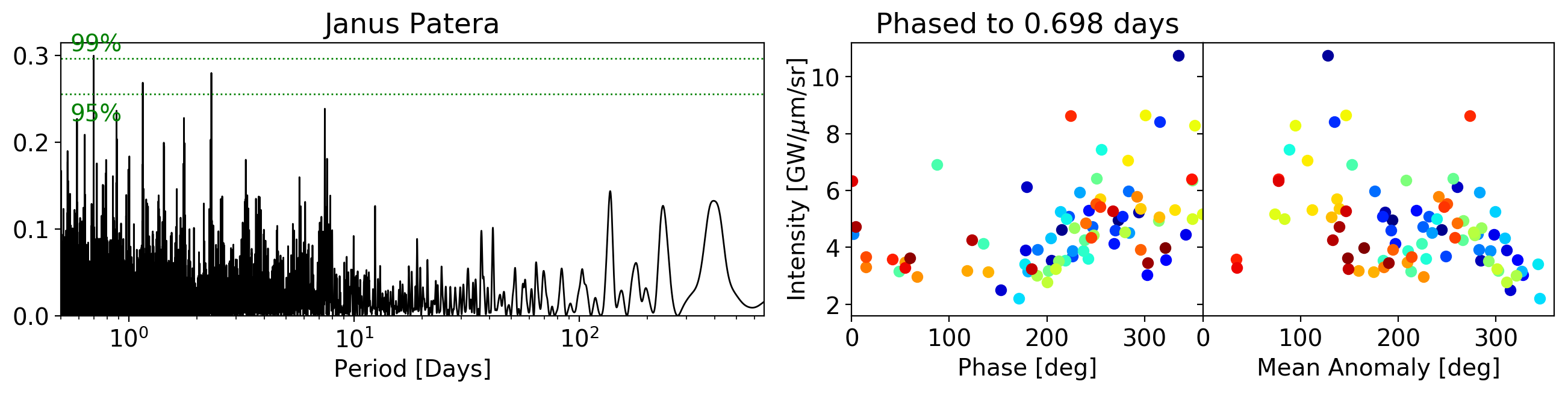}
\includegraphics[width=16cm]{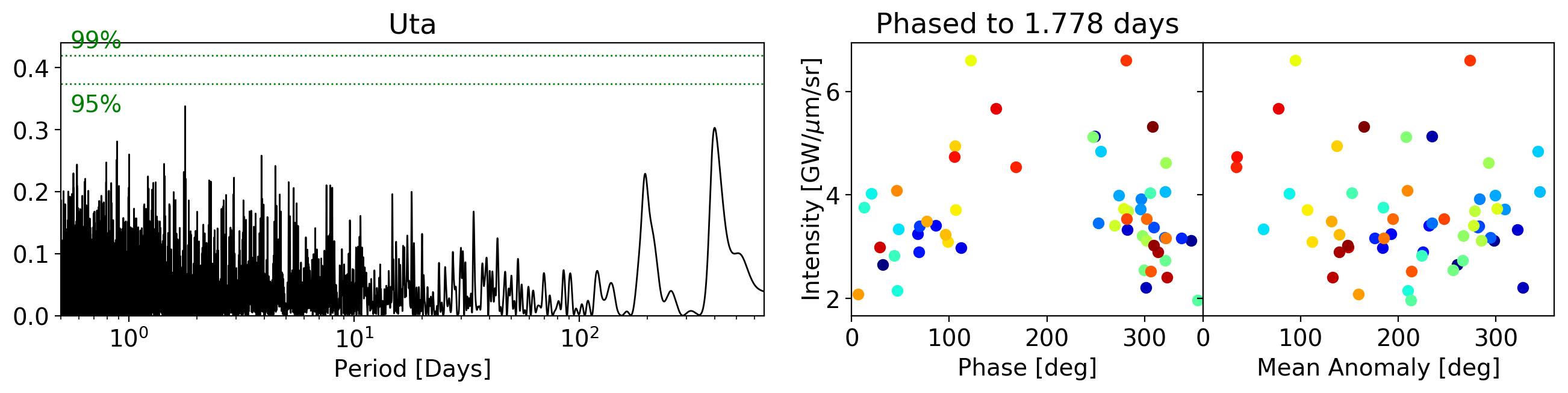}
\includegraphics[width=16cm]{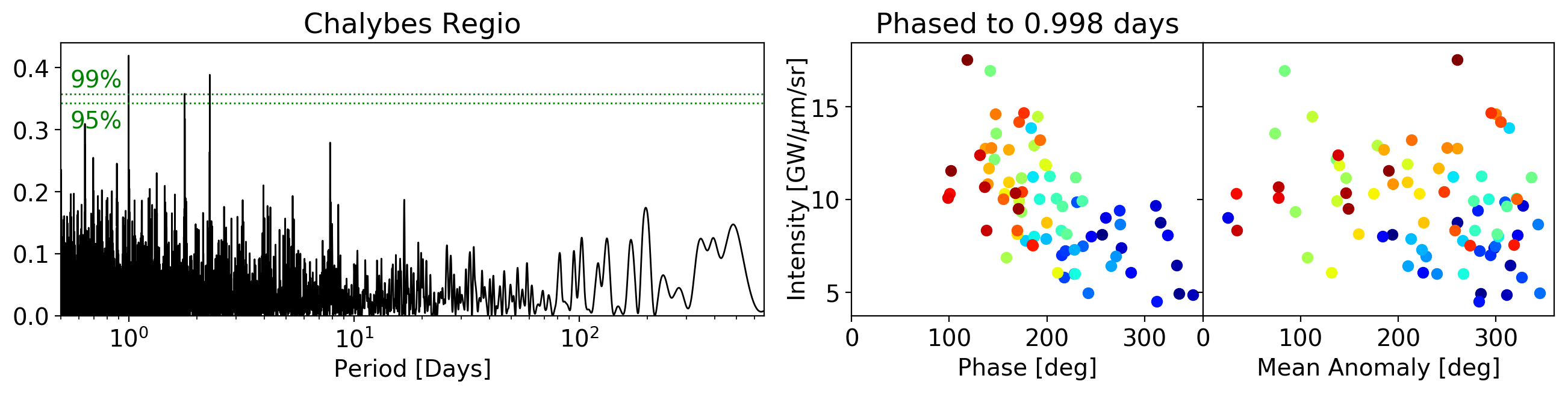}
\includegraphics[width=16cm]{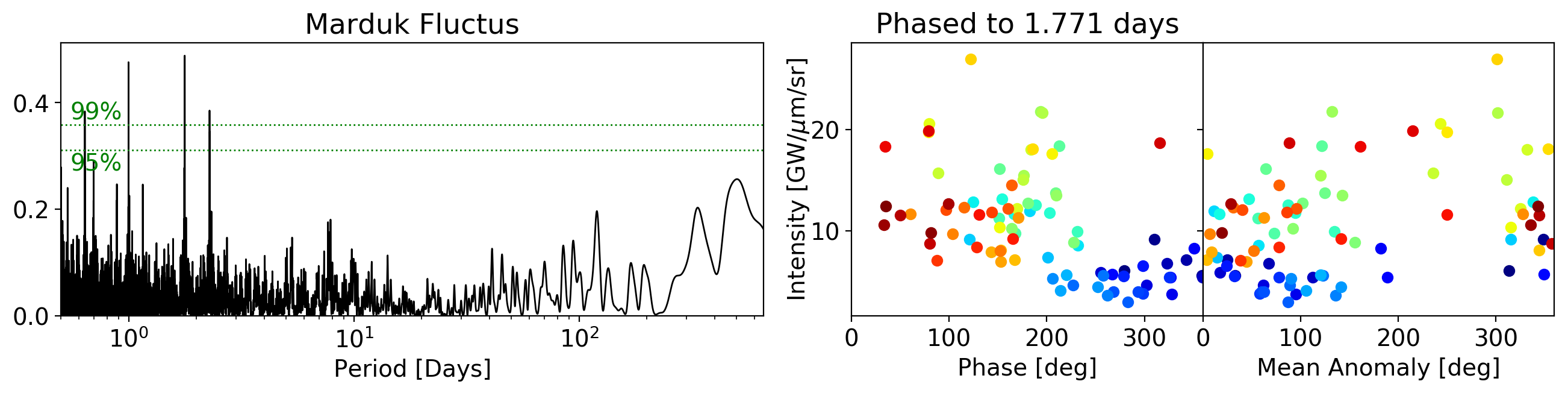}
\includegraphics[width=16cm]{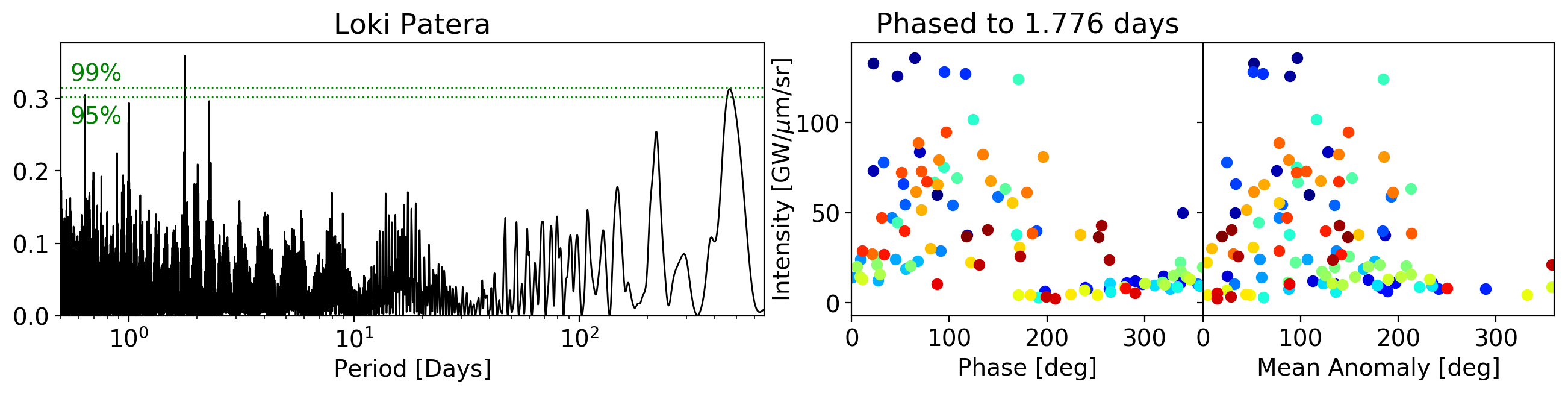}
\caption{Generalized Lomb-Scargle periodograms for the five most consistently detected hot spots. The 99\% and 95\% significance levels are shown as dotted horizontal lines. The plots in the middle column show the data phased to the period corresponding to the peak in the periodogram, and the rightmost column shows the data as a function of Io's mean anomaly at the time of observation. The most prominent periodicities are near Io's and Earth's rotation periods (1 and 1.77 days), and their beat frequencies (periods near 0.6 and 2.3 days). The mean anomaly plots demonstrate that the 1.77-day periodicities are more consistent with being an observing cadence effect rather than a physical effect due to tidally modulated volcanism, which would exhibit a shorter period and a mean anomaly correlation. Within a given plot, the coloring of points is monotonic with time (blue=earliest; red=latest) to indicate the temporal ordering of the datapoints.\label{fig:LS}}
\end{figure}

We conducted a periodicity analysis on the five persistently active hot spots listed above by calculating Lomb-Scargle periodograms (Scargle 1982; Zechmeister and K\"urster 2009) and comparing the periodogram peaks against significance levels derived by bootstrapping. These periodograms are shown, with significance levels indicated, in Figure \ref{fig:LS}, and the volcano intensity timeline is plotted phased on the period corresponding to the periodogram peak. The bootstrapping technique samples from the dataset randomly with replacement and computes the periodogram; the confidence levels correspond to the percentage of resampled datasets that show no peaks above the indicated level (Ivezi\'c et al. 2014; VanderPlas 2018). Note that because the duration of Io's eruptions is typically longer than the interval between observations, confidence intervals derived from random resampling will lead to an apparent enhancement in the significance of observing cadence periodicities. \par
Nearly all hot spots show peaks near 0.997 days and 1.77 days, with weaker signals near 0.64 and 2.3 days (seen prominently in many of the periodograms in Figure \ref{fig:LS}), which correspond to Earth's sidereal day, Io's rotation period (sidereal period = 1.7691 days), and the periods corresponding to their beat frequencies ($\nu_{beat}=\mid \nu_1-\nu_2\mid$). These periods reflect the observing cadence: the average interval between observations is a multiple of Earth's sidereal day, while repeat observations of a given hot spot are made (on average) at multiples of Io's rotation period. Io's rotation period as observed from Earth differs slightly from its sidereal period due to the relative motion of Earth and Jupiter, and is minimized at opposition when the motion of Earth relative to Jupiter is maximized perpendicular to the line of sight. This leads to Earth-apparent rotation periods in the range of 1.7680-1.7691 days. \par
A periodicity at Io's rotation period could also be an indication of a tidally modulated volcanic process, whereby the volcanic activity or thermal emission is controlled in part by diurnally varying tidal stresses. However, the periodicities near Io's rotation period in the dataset analyzed here do not show evidence for this effect. In particular, the peak periodogram power is at periods of 1.769-1.776 days, which match or slightly exceed Io's apparent rotation period but are a poorer match to Io's 1.7627 day anomalistic period, or time between successive perijoves, which is the relevant parameter for diurnal stresses and differs from the apparent rotation period due to the precession of Io's orbit. In order to further highlight the difference between the observed 1.77-day signal and Io's tidal forcing, the rightmost column in Figure \ref{fig:LS} shows the brightness of each volcano as a function of Io's mean anomaly at the time of observation, demonstrating that there is no mean anomaly correlation even in hot spots that show a strong 1.77-day periodicity. \par
The period of precession of Io's longitude of perijove is $\sim$1.5 years, so that on the timescale of a few months we see the same Ionian longitudes near a similar phase of Io's orbit, which likely accounts for the prominence of both Earth's and Io's rotation periods in the periodograms. This can be seen in the middle column of Figure \ref{fig:LS}: all datapoints are color-coded by time (cycling from blue to red from early to late), and it is clear from the middle plot of the Chalybes Regio panel, for example, that the apparent periodicity likely arises from a combination of a long-term brightening and observing cadence biases. In essence, we are unable to rigorously distinguish between a scenario where a volcano is variable over an orbital period, and a scenario where it is variable over a longer timescale but observational effects caused apparent orbital-timescale periodicities. \par
This limitation could be entirely eliminated in a spacecraft dataset, where the observing cadence is less regular and the same hot spot can be viewed at a variety of mean anomalies within a time period that is short relative to the timescale for intrinsic variability of that volcano. \par
Only Loki Patera shows a statistically significant periodicity at a period other than the four discussed above. Over the time period of observation analyzed here, Loki Patera's activity was periodic with a period of 465.63 days.
\subsection{Hot spot spatial distribution} \label{sec:spatdist}
\begin{figure}
\centering
\includegraphics[width=14cm]{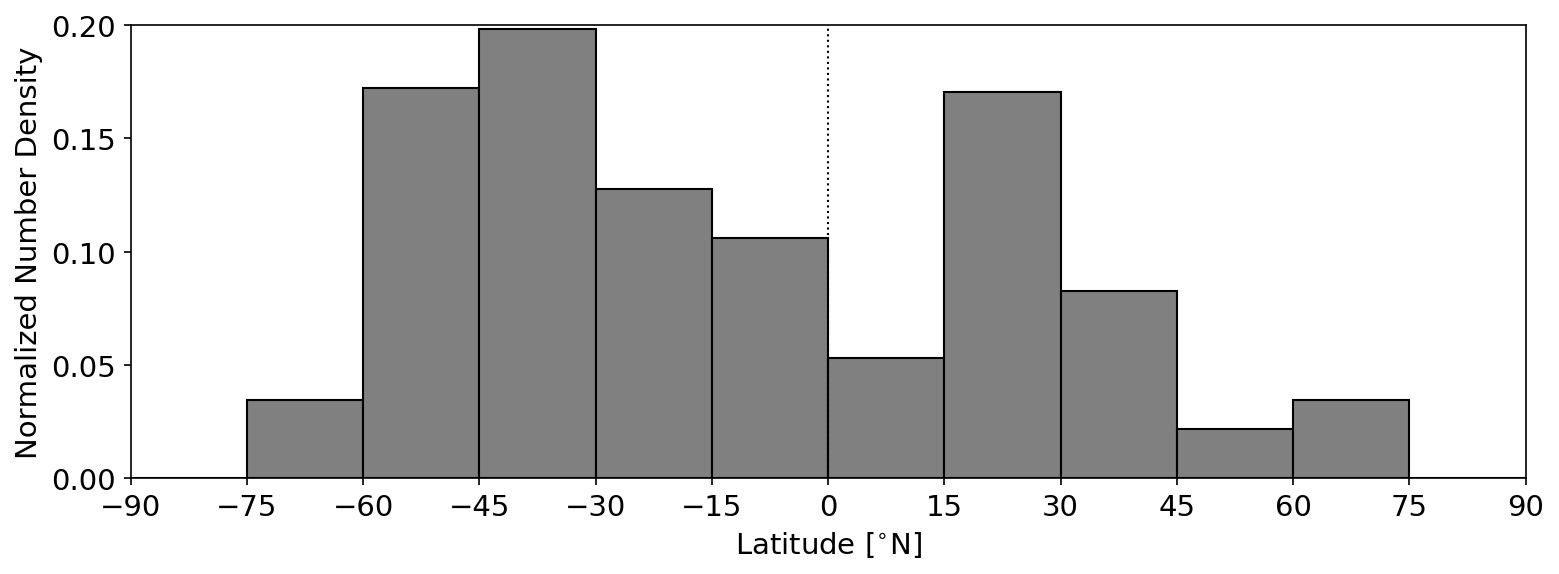}
\includegraphics[width=14cm]{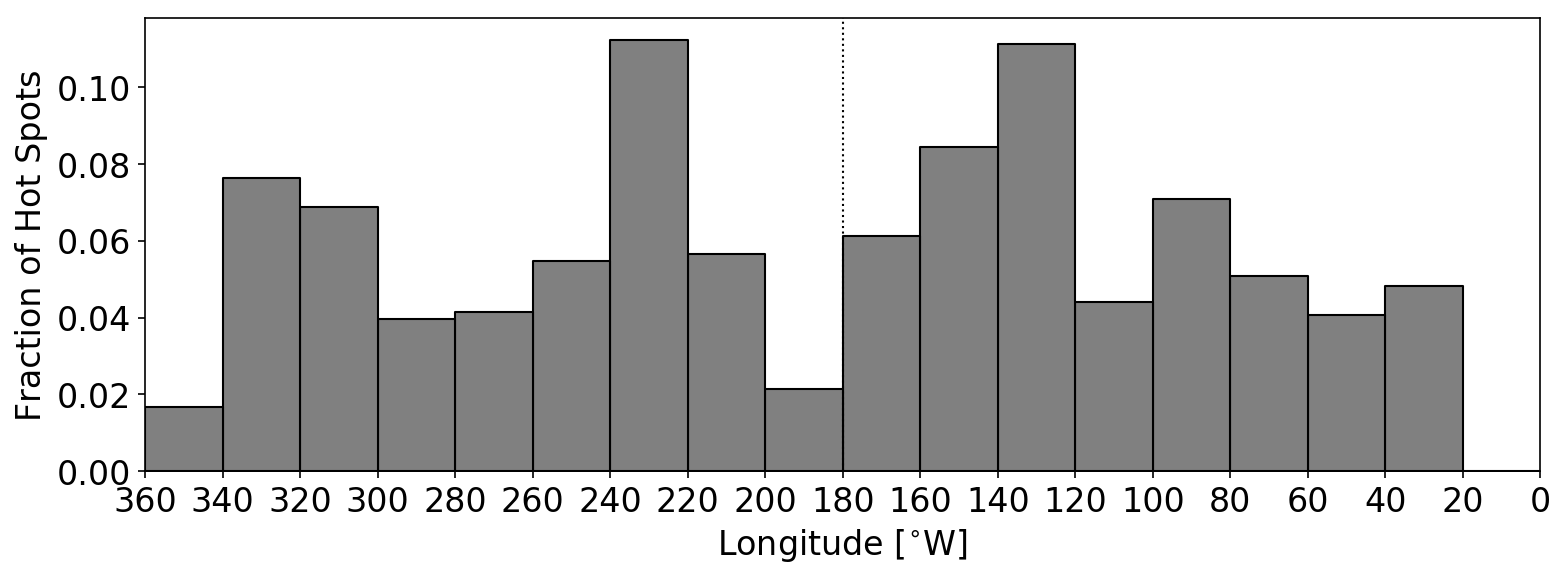}
\caption{Distribution of detected hot spots in latitude and longitude. The longitude distribution is plotted as fraction of total hot spot number per longitude bin, using only hot spots that were detected at L', and is corrected for observational biases. The latitude distribution is corrected for the surface area in each latitude bin and normalized but is not corrected for observational biases, which contribute to the dearth of hot spots at high latitudes. \label{fig:LatLonHists}}
\end{figure}

\begin{figure}
\centering
\includegraphics[width=14cm]{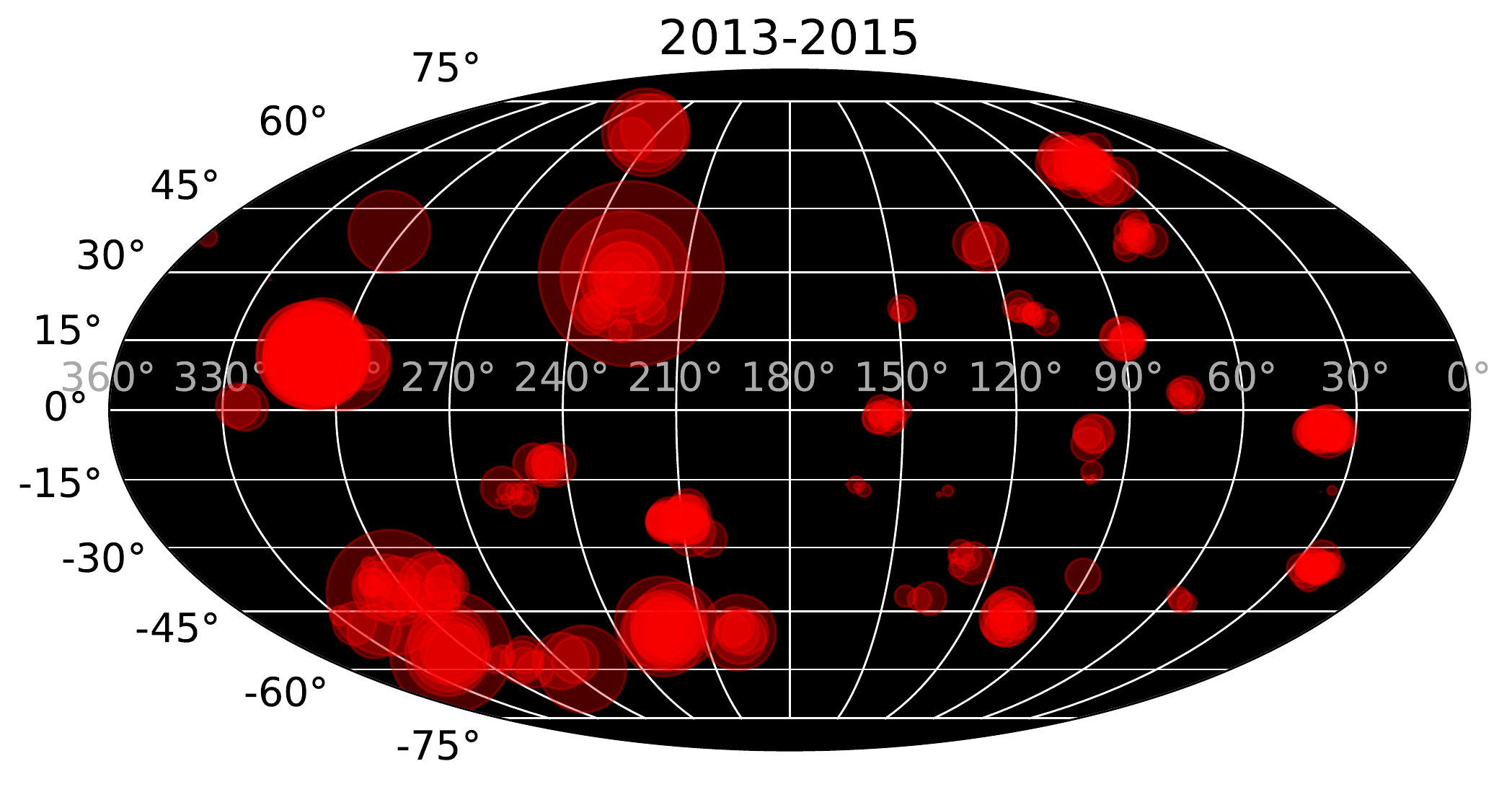}
\includegraphics[width=14cm]{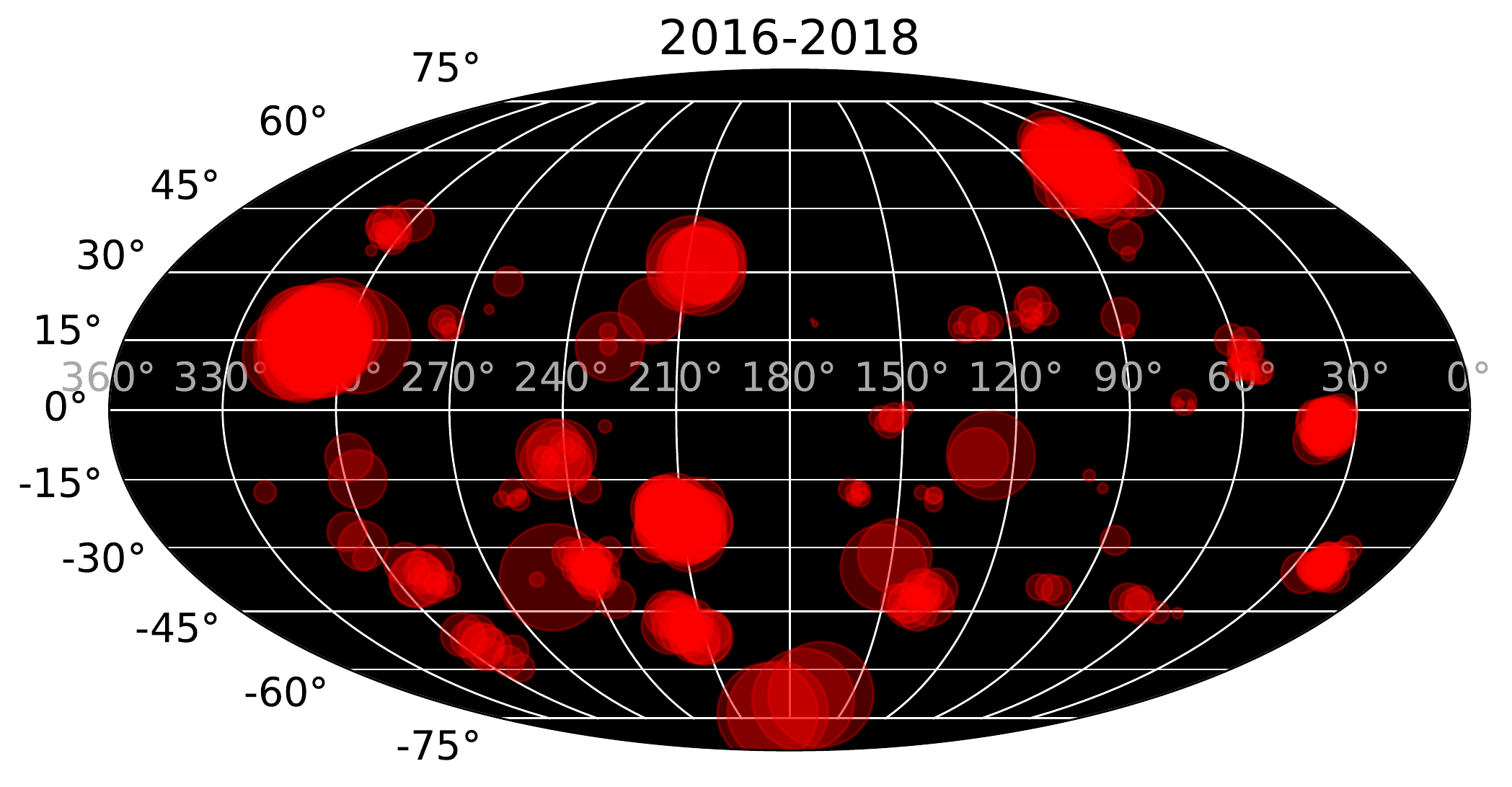}
\includegraphics[width=14cm]{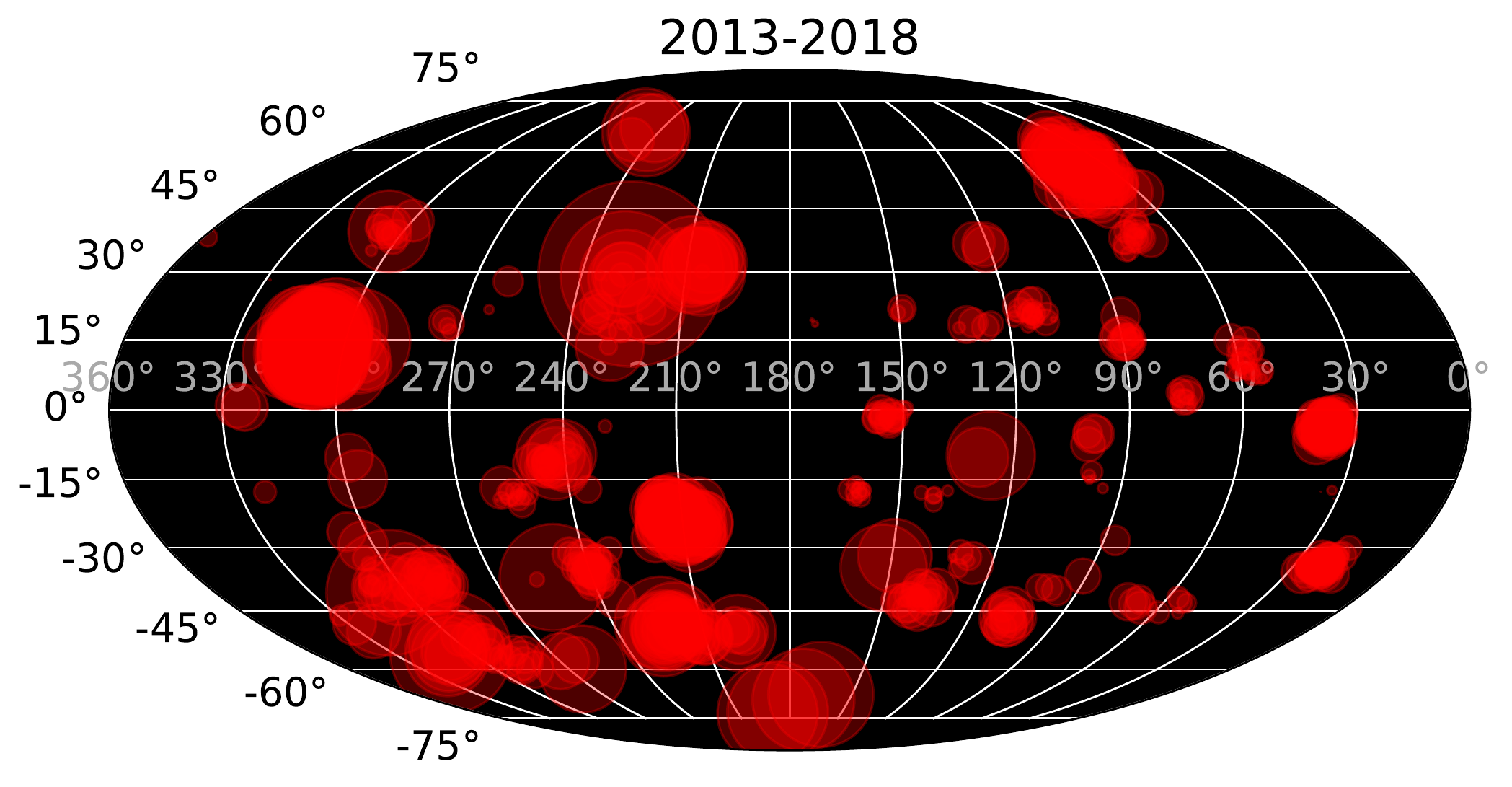}
\caption{The spatial distribution of hot spot thermal emission detected on Io in 2013-2018. Each circle shows the location and brightness of a single hot spot detection, with circle size proportional to the log of the 3.8-$\mu$m intensity. All circles are semi-transparent, and high-opacity regions indicate multiple detections at the same location. The top and middle panels show the distribution in two distinct time periods (the 2013-2015 plot is identical to that given in de Kleer and de Pater 2016a), and the bottom panel shows the cumulative distribution from 2013-2018. The position uncertainties are typically a few degrees at low latitudes and higher towards the poles; much but not all of the apparent jitter in hot spot locations is within these uncertainties.\label{fig:spatdist}}
\end{figure}

The distribution of hot spot number density with latitude and longitude is shown in Figure \ref{fig:LatLonHists}, and Figure \ref{fig:spatdist} plots the location and brightness of all L'-band hot spot detections, updated from a similar figure based on the 2013-2015 data in de Kleer and de Pater (2016b). The longitudinal distribution has been corrected for the sensitivity of the observations to each longitude bin. The latitudinal distribution is not corrected because the latitudinal differences in the volcano brightness distribution or in topography are poorly constrained. \par
Our observations in 2013-2015 showed an apparent difference in the spatial distribution of bright, transient eruptions compared to persistent hot spots (de Kleer and de Pater 2016b). In particular, all nine of the volcanoes that hosted bright eruptions ($I_{max,Lp}>$30 GW/$\mu$m/sr) during those years are located on the trailing hemisphere. In the full dataset (2013-2018), 18 volcanoes exhibited bright eruptions, where bright is defined as $I_{max,Lp}>$20 GW/$\mu$m/sr. Note that this definition is effectively the same as in our previous paper because there were no volcanoes with detected $I_{max,Lp}$ between 20 and 30 GW/$\mu$m/sr in 2013-2015. The threshold was lowered because the larger dataset now available indicates that no volcano persistently maintains a flux density level above 20 GW/$\mu$m/sr, and this cutoff is therefore sufficient to isolate bright transient events. Of these 18 volcanoes, every single one falls within a 180$^{\circ}$ band in longitude, from 128-308$^{\circ}$W, despite the fact that our program had comparable sensitivity to events of this magnitude at all Ionian longitudes. Moreover, all but two of the eruptions occurred on Io's trailing hemisphere (180-360$^{\circ}$W); the probability of 16 or more eruptions occurring on the trailing hemisphere (given 18 eruptions total) is 0.00066 if volcanoes are randomly distributed in longitude. \par
Despite the distinctive distribution of the largest eruptions, there is no significant difference between the two hemispheres in terms of spatially and temporally averaged near-infrared brightness (provided that Loki Patera is excluded), nor in the number of active hot spots. The time-averaged volcanic L'-band intensity arises 47.5\% and 52.5\% from the leading and trailing hemispheres, respectively (excluding Loki Patera), while the number of hot spots detected at L' is identical between the two hemispheres (31 hot spots, excluding those detected only at longer wavelengths, to which only a subset of the data were sensitive). \par 
In order to further explore whether any hemispheric-scale asymmetries are present in the distribution of the hot spots, we broaden this analysis from a comparison of just leading vs. trailing hemispheres by choosing all 180-degree longitude intervals and determining the fraction of hot spots that fall within each interval. 
While the leading and trailing hemispheres exhibit comparable time-averaged radiances and hot spot number, there is an asymmetry between the sub- and anti-jovian hemispheres with more hot spots and higher radiances on the anti-jovian hemisphere, despite the fact that this hemisphere had poorer coverage during our program. The hemisphere centered on 160$^{\circ}$W maximizes both metrics, containing $\sim$60\% of the hot spots and $>$70\% of the time averaged radiances. However, in artificial datasets where hot spots are randomly distributed in longitude, an asymmetry in hot spot number at this level is well within the expected range (i.e. within one $\sigma$ of the median). \par
\section{Conclusions} \label{sec:conc}
We present results from measurements of the thermal emission of Io's volcanoes, derived from near-infrared imaging with adaptive optics at the Keck and Gemini N telescopes on 271 nights between August 2013 and the end of 2018. The first 100 nights of observations were presented in de Kleer and de Pater (2016a), while the 171 nights since the start of 2016 are presented here for the first time. Over the five years of the program to date, we made 980 detections of over 75 unique hot spots, with some hot spots detected more than 80 times and Loki Patera detected 113 times. We provide downloadable tables of hot spot brightnesses and observing details, and hope that these data products will serve as a resource for others in the community who will build on the analyses presented here.\par
Nearly all bright transient eruptions where temperature measurements were possible displayed temperatures above 800 K, confirming that eruptions at such high temperatures are common and are likely the rule rather than the exception. The detection of new hot spots that were not previously detected by spacecraft is a common occurrence. Adding the data presented here to that summarized by Cantrall et al. (2018), there have now been 104 distinct hot spots seen in the AO data from 2001-2018, 25-30 of which were not previously seen by spacecraft. It is likely that many of these hot spots have been active since before the \textit{Galileo} and \textit{Voyager} visits but were not emitting sufficient radiation during the visits to have been detected. However, some of the new hot spots have no corresponding surface feature and remain persistently active after they are first detected (e.g. Chalybes Regio and the hot spot SE of Pele), suggesting that activity recently initiated at these locations. We performed a periodicity search on the five most consistently detected hot spots (each detected 57-113 times) but did not detect any new periodicities beyond those introduced by the observing cadence. Spacecraft data would be needed to draw a robust conclusion about tidally modulated volcanism on diurnal timescales.\par
De Kleer and de Pater (2016b) noted that all bright, transient eruptions took place on Io's trailing hemisphere. This trend continues through the additional 3 years of data presented here, and the probability of the observed asymmetry is 0.00066 if volcanoes are randomly distributed. Note that this asymmetry applies only to the character of the volcanism; the number and cumulative near-IR radiance is nearly identical between leading and trailing hemispheres. \par
This dataset now constitutes the largest set of unique detections of thermal emission from individual Ionian hot spots to date, permitting robust statistical analyses of properties such as the spatial distribution of hot spot activity, the variability and time-averaged power of numerous individual hot spots, and the occurrence rates of bright and/or high-temperature eruptions. These data, in combination with \textit{Galileo's} sensitivity to smaller, cooler hot spots and the multi-decadal time baseline provided by ground-based occultation data, are now providing a truly global, multi-wavelength picture of Io's volcanic activity over a wide range of timescales. \par
The timing of our program coincided with intensive observations of the extended sodium cloud and the plasma torus by ground-based programs and by the \textit{EXCEED/Hisaki} and \textit{Juno} missions. The correlation of these datasets with our timeline of Io's activity is already providing clues into the connections between different components of the jovian system (Yoshikawa et al. 2017; Koga et al. 2018; Morgenthaler et al. 2019), but our understanding of this system is far from complete. Continued coverage of Io's volcanoes throughout these missions will be key to unraveling the sources of variability in the jovian neutral and plasma environment.
\section*{Acknowledgements}
KdK is supported by the Heising-Simons Foundation through a \textit{51 Pegasi b} postdoctoral fellowship, and this research was partially supported by the National Science Foundation grant AST-1313485 to UC Berkeley and a NASA Keck PI Data Award, administered by the NASA Exoplanet Science Institute. We are grateful to Roy and Frances Simperman for their support of the Keck Visiting Scholars program, which enabled KdK, EM, and CA to develop the Keck twilight program through which some of the data presented here were obtained. We thank G. Puniwai for acquiring several of the Keck observations. We thank P. Capak, J. Cohen, N. Hernitschek, D. Masters, and S.A. Stanford of the the Complete Calibration of the Color-Redshift Relation (C3R2; Masters et al. 2017) NASA Keck Key Strategic Mission Support survey team for providing twilight observations on the nights of UT 2017 December 11-13. The work of DS and AGD was carried out at the Jet Propulsion Laboratory, California Institute of Technology, under a contract with NASA. Much of the data presented herein were obtained at the Gemini Observatory, which is operated by the Association of Universities for Research in Astronomy, Inc., under a cooperative agreement with the NSF on behalf of the Gemini partnership: the National Science Foundation (United States), National Research Council (Canada), CONICYT (Chile), Ministerio de Ciencia, Tecnolog\'{i}a e Innovaci\'{o}n Productiva (Argentina), Minist\'{e}rio da Ci\^{e}ncia, Tecnologia e Inova\c{c}\~{a}o (Brazil), and Korea Astronomy and Space Science Institute (Republic of Korea). Some of the data presented herein were obtained at the W. M. Keck Observatory, which is operated as a scientific partnership among the California Institute of Technology, the University of California and the National Aeronautics and Space Administration. Some of the data was obtained at the W. M. Keck Observatory from telescope time allocated to the National Aeronautics and Space Administration through the agency's scientific partnership with the California Institute of Technology and the University of California. The Observatory was made possible by the generous financial support of the W. M. Keck Foundation. The authors wish to recognize and acknowledge the very significant cultural role and reverence that the summit of Maunakea has always had within the indigenous Hawaiian community.  We are most fortunate to have the opportunity to conduct observations from this mountain.
\clearpage
\startlongtable
\begin{deluxetable}{llllll}
\tabletypesize{\small}
\tablecaption{Overview of hot spots\label{tbl:overview}}
\tablehead{Site & Lat & Lon & N$_{det}$ & $\overline{F}_{filt}$$^a$ & Filter$^a$ \\
 & $^{\circ}$N & $^{\circ}$W & & GW/$\mu$m/sr & }
\startdata
Nusku Patera & -65.0 & 6.2 & 1 & \multicolumn{2}{c}{Narrowband Only} \\
Uta & -34.4 & 21.0 & 57 & 3.6 & Lp \\
Kanehekili Fluctus & -17.0 & 34.5 & 8 & 1.2 & Lp \\
Janus Patera & -3.9 & 37.4 & 84 & 4.7 & Lp \\
UP 38W & -25.3 & 37.7 & 1 & 1.9 & Ms \\
Pfu374 & -24.3 & 49.7 & 3 & 1.0 & Ms \\
Masubi & -42.9 & 53.7 & 9 & 2.4 & Lp \\
PFd1691 & 9.4 & 58.3 & 22 & 2.5 & Lp \\
Laki-Oi Patera & -44.6 & 59.7 & 4 & 3.9 & Lp \\
Shamshu Patera & -8.3 & 61.5 & 1 & 1.1 & Ms \\
Tejeto Patera & -42.9 & 68.7 & 4 & 4.4 & Lp \\
Chalybes Regio & 55.4 & 70.2 & 80 & 9.6 & Lp \\
Zal Patera & 37.9 & 74.6 & 24 & 2.9 & Lp \\
Tawhaki Patera & 2.5 & 75.6 & 19 & 2.1 & Lp \\
Ekhi Patera & -28.4 & 86.7 & 1 & 3.8 & Lp \\
Gish Bar & 15.6 & 89.1 & 18 & 3.4 & Lp \\
Aluna Patera & 41.7 & 90.1 & 2 & 3.2 & Ms \\
P207 & -36.5 & 91.1 & 1 & 5.0 & Lp \\
Shango Patera & 33.5 & 95.6 & 3 & 1.8 & Ms \\
Itzamna Patera & -15.0 & 99.0 & 9 & 1.6 & Lp \\
Arusha Patera & -39.6 & 99.0 & 4 & 3.5 & Lp \\
Sigurd Patera & -5.1 & 99.2 & 8 & 4.4 & Lp \\
P197 & -46.9 & 107.3 & 11 & 6.2 & Lp \\
Amirani & 20.5 & 113.2 & 27 & 2.6 & Lp \\
Dusura Patera & 36.4 & 121.1 & 3 & 7.5 & Lp \\
Maui Patera & 18.2 & 125.8 & 2 & 3.2 & Lp \\
P95 & -10.0 & 127.8 & 2 & 36.5 & Lp \\
Malik Patera & -32.9 & 129.6 & 9 & 3.1 & Lp \\
UP 132W & 18.4 & 131.6 & 5 & 3.6 & Lp \\
Thor & 40.6 & 134.7 & 2 & 1.3 & Ms \\
P123 & -41.9 & 139.2 & 20 & 4.6 & Lp \\
Tupan Patera & -18.0 & 140.5 & 10 & 1.6 & Lp \\
Surya Patera & 21.2 & 149.4 & 4 & 2.7 & Lp \\
Shamash Patera & -33.2 & 150.5 & 2 & 41.3 & Lp \\
Sobo Fluctus & 12.9 & 152.8 & 1 & \multicolumn{2}{c}{Narrowband Only} \\
Prometheus & -1.5 & 153.3 & 22 & 2.8 & Lp \\
Culann & -17.2 & 161.8 & 11 & 2.0 & Lp \\
Zamama & 18.5 & 173.2 & 3 & 1.2 & Lp \\
Illyrikon Regio & -70.8 & 179.9 & 4 & 109.2 & Lp \\
Sethlaus/Gabija Paterae & -50.0 & 198.1 & 6 & 11.6 & Lp \\
Isum Patera & 31.1 & 205.4 & 16 & 37.6 & Lp \\
Marduk Fluctus & -23.7 & 211.1 & 87 & 10.5 & Lp \\
Kurdalagon & -49.3 & 216.7 & 36 & 13.1 & Lp \\
Unknown & 53.6 & 217.8 & 1 & \multicolumn{2}{c}{Narrowband Only} \\
Susanoo/Mulungu Paterae & 18.6 & 221.0 & 10 & 4.5 & Lp \\
201308C & 29.1 & 228.0 & 11 & 555.7 & Lp \\
P17 & -3.5 & 228.8 & 1 & 1.8 & Lp \\
P13 & 13.9 & 229.0 & 4 & 9.2 & Lp \\
East Girru & 21.3 & 233.5 & 3 & 4.9 & Lp \\
Reiden Patera & -18.0 & 234.4 & 2 & 3.5 & Lp \\
Pyerun Patera & -57.7 & 237.1 & 1 & 3.9 & Ms \\
SE of Pele & -34.5 & 239.5 & 30 & 3.9 & Lp \\
Pillan Patera & -11.3 & 243.7 & 21 & 7.1 & Lp \\
Chors Patera & 65.1 & 245.6 & 5 & 30.6 & Lp \\
UP 254W & -37.1 & 254.5 & 2 & 67.7 & Lp \\
Pele & -18.2 & 255.2 & 19 & 2.2 & Lp \\
Shakuru Patera & 24.8 & 261.7 & 2 & 2.7 & Lp \\
Mithra Patera & -58.0 & 265.6 & 4 & 25.4 & Lp \\
Svarog Patera & -51.6 & 269.3 & 3 & 4.1 & Ms \\
Daedalus Patera & 18.7 & 273.9 & 5 & 2.5 & Lp \\
PV59 & -38.2 & 289.7 & 22 & 6.7 & Lp \\
N Lerna Regio & -56.0 & 290.6 & 19 & 5.2 & Lp \\
Kibero Patera & -12.5 & 297.1 & 2 & 11.7 & Lp \\
Amaterasu Patera & 38.8 & 304.3 & 13 & 7.6 & Lp \\
Sengen Patera & -29.8 & 305.1 & 4 & 5.2 & Lp \\
Rarog Patera & -39.2 & 305.4 & 14 & 29.3 & Lp \\
Heno Patera & -55.6 & 307.5 & 7 & 70.3 & Lp \\
Loki Patera & 12.6 & 307.5 & 113 & 38.3 & Lp \\
Shoshu Patera & -17.6 & 322.9 & 1 & 2.7 & Lp \\
Tol-Ava Patera & 0.7 & 326.5 & 4 & 4.4 & Lp \\
PV170 & -47.9 & 327.8 & 3 & 7.1 & Lp \\
Fuchi Patera & 28.3 & 328.7 & 1 & 1.0 & Lp \\
Surt & 44.4 & 334.1 & 2 & 1.2 & Ms \\
Pfu1063 & 41.7 & 357.7 & 3 & 1.7 & Lp \\
Paive Patera & -42.9 & 358.3 & 2 & 0.9 & Ms \\
\enddata
\tablenotetext{a}{The mean observed flux density $\overline{F}_{filt}$ is given in the Lp filter if data were available in this filter, and in the Ms filter if no Lp detections were made. If no detections were made in either broadband filter, detection in narrowband only is indicated.}
\end{deluxetable}
\begin{table}
\caption{Bright eruptions$^a$, 2013-2018 \label{tbl:transients}}
\begin{center}
\begin{tabular}{lllllc}
\hline
Site & Date of Peak & Lat & Lon & I$_{max,Lp}$ & Reference \\
 & [UT] & [$^{\circ}$N] & [$^{\circ}$W] & [GW/$\mu$m/sr] & \\
 \hline
Heno Patera & 08-15-2013 & -56 & 308 & 270$\pm$70 & c \\
Rarog Patera & 08-15-2013 & -39 & 305 & 325$\pm$80 & c \\
Loki Patera$^b$ & 08-22-2013 & 13 & 308 & 136$\pm$20 & c \\
201308C & 08-29-2013 & 29 & 228 & $>$500 & d \\
Chors Patera & 10-22-2014 & 65 & 246 & 57$\pm$19 & e \\
Mithra Patera & 01-10-2015 & -58 & 266 & 55$\pm$12 & e \\
Sethlaus/Gabija Paterae & 04-01-2015 & -50 & 198 & 33$\pm$5 & e \\
Kurdalagon$^b$ & 04-05-2015 & -49 & 217 & 68$\pm$11 & e \\
Amaterasu Patera & 12-25-2015 & 39 & 304 & 43$\pm$6 & e \\
P95 & 05-17-2016 & -10 & 128 & 58$\pm$13\\
Shamash Patera & 06-20-2016 & -33 & 151 & 53$\pm$9\\
Illyrikon Regio & 06-27-2016 & -71 & 180 & 125$\pm$69\\
P13 & 02-05-2017 & 14 & 229 & 23$\pm$2\\
Marduk Fluctus$^b$ & 02-05-2017 & -24 & 211 & 27$\pm$2\\
Pillan Patera$^b$ & 02-23-2017 & -11 & 244 & 27$\pm$5\\
Susanoo/Mulungu Paterae & 01-12-2018 & 19 & 221 & 20$\pm$3\\
UP 254W & 05-10-2018 & -37 & 252 & 134$\pm$24\\
Isum Patera & 05-27-2018 & 31 & 205 & 64$\pm$16\\
\end{tabular}\\
\end{center}
$^a$All eruptions detected with $I_{max,Lp}>$20 GW/$\mu$m/sr during this time period.\\
$^b$Nearly all bright eruptions were transient events at sites where activity was not otherwise detected. Exceptions are: Pillan Patera, Loki Patera, and Marduk Fluctus, which were persistently active but exhibited spikes in activity; and Kurdalagon Patera, which was not detected prior to its first eruption but remained detectable afterwards.\\
$^c$de Pater et al. (2014)\\
$^d$de Kleer et al. (2014)\\
$^e$de Kleer and de Pater (2016a)
\end{table}
\begin{table}
\caption{High-temperature eruptions$^a$, 2013-2018 \label{tbl:highT}}
\begin{center}
\begin{tabular}{llllc}
\hline
Site & Date & $\mu$ & T$^b$ & Reference \\
 & [UT] & & [K] & \\
 \hline
Shamash Patera & 2016-Jun-20 & 0.81 & 1000$\pm$110 \\
& 2016-Jun-27 & 0.74 & 850$\pm$80 \\
Culann & 2017-Jun-16 & 0.81 & 860$\pm$140 \\
UP 254W & 2018-May-10 & 0.63 & 960$\pm$100 \\
PV170 & 2014-Dec-02 & 0.42 & 850$\pm$40 & c \\
Isum Patera & 2018-May-27 & 0.48 & 1200$\pm$220 \\
& 2018-May-31 & 0.70 & 1180$\pm$120 \\
& 2018-Jun-16 & 0.84 & 1120$\pm$100 \\
& 2018-Jun-18 & 0.43 & 1440$\pm$410 \\
& 2018-Jun-23 & 0.85 & 980$\pm$70 \\
& 2018-Jun-25 & 0.62 & 1230$\pm$280 \\
& 2018-Jun-30 & 0.85 & 1010$\pm$80 \\
PFd1691 & 2018-Jan-19 & 0.98 & 830$\pm$80 \\
Rarog Patera & 2013-Aug-15 & 0.60 &1300$\pm$200 & d \\
 & 2014-Feb-10 & 0.78 & 890$\pm$120 & c \\
 & 2015-Mar-31 & 0.76 & 950$\pm$60 & c \\
PV59 & 2014-Oct-31 & 0.59 & 950$\pm$200 & c \\
P95 & 2016-May-17 & 0.39 & 1020$\pm$180 \\
Kurdalagon Patera & 2015-Jan-26 & 0.54 & 1200$\pm$150 & c \\
 & 2015-Mar-31 & 0.26 & 820$\pm$110 & c \\
 & 2015-Apr-05 & 0.57 & 1300$\pm$200 & c \\
Tawhaki Patera & 2014-Mar-11 & 0.54 & 900$\pm$170 & c \\
 & 2018-Jan-19 & 0.97 & 800$\pm$90 \\
P197 & 2014-Mar-11 & 0.67 & 1000$\pm$250 & c \\
N Lerna Regio & 2014-Dec-02 & 0.50 & 820$\pm$180 & c \\
 & 2015-Mar-31 & 0.54 & 940$\pm$120 & c \\
Reiden Patera & 2017-Dec-12 & 0.90 & 1170$\pm$100 \\
201308C & 2014-Dec-02 & 0.64 & 850$\pm$160 & e \\
SE of Pele & 2017-Dec-12 & 0.79 & 950$\pm$160 \\
P123 & 2015-Jan-11 & 0.74 & 820$\pm$160 & c \\
Illyrikon Regio & 2016-Jun-20 & 0.24 & 1210$\pm$690 \\
& 2016-Jun-27 & 0.17 & 1060$\pm$340 \\\end{tabular}
\end{center}
$^a$All eruptions detected with T$>$800 K during this period.\\
$^b$Temperatures are derived from intensities corrected for geometric foreshortening, and may be overestimated in observations with high emission angle ($\mu$) if fire fountaining is producing a substantial fraction of the short-wavelength emission.\\
$^c$de Kleer and de Pater (2016a)\\
$^d$de Pater et al. (2014)\\
$^e$de Kleer et al. (2014)
\end{table}
%\clearpage
\clearpage
\section*{References}
\begin{itemize}
\item[]{Cantrall, C., de Kleer, K., de Pater, I., et al. Variability and geologic associations of volcanic activity on Io in 2001-2016. Icarus 312, 267-294 (2018).}
\item[]{Carlson, R.W., Weissman, P.R., Smythe, W.D., Mahoney, J.C. Near-Infrared Mapping Spectrometer experiment on Galileo. \textit{Space Sci Rev} 60, 457-502 (1992).}
\item[]{Carr, M.H. Silicate volcanism on Io. JGR 91, 3521-3532 (1986).}
\item[]{Davies, A.G. Volcanism on Io: a Comparison with Earth, Cam. Univ. Press (2007).}
\item[]{Davies, A.G., Keszthelyi, L.P., Harris, A.J.L. The thermal signature of volcanic eruptions on Io and Earth. J. Volc. Geo. Res. 194, 75-99 (2010).}
\item[]{Davies, A.G., Veeder, G.J., Matson, D.L., Johnson, T.V. Io: Charting thermal emission variability with the Galileo NIMS Io Thermal Emission Database (NITED): Loki Patera. GeoRL 39, L01201, p. 1-6 (2012).}
\item[]{Davies, A.G., Davies, R.L., Veeder, G.J., et al. Discovery of a powerful, transient, explosive thermal event at Marduk Fluctus, Io, in \textit{Galileo} NIMS data. \textit{GRL} 45, 2926-2933 (2018).} 
\item[]{de Kleer, K., de Pater, I., Davies, A.G., et al. Near-infrared monitoring of Io and detection of a violent outburst on 29 August 2013. Icarus 242, 352-364 (2014).}
\item[]{de Kleer, K., de Pater, I. Time variability of Io's volcanic activity from near-IR adaptive optics observations on 100 nights in 2013-2015. Icarus 280, 378-404 (2016a).}
\item[]{de Kleer, K., de Pater, I. Spatial distribution of Io's volcanic activity from near-IR adaptive optics observations on 100 nights in 2013-2015. Icarus 280, 405-414 (2016b).}
\item[]{de Pater, I., Davies, A.G., \'Ad\'amkovics, M., Ciardi, D.R. Two new, rare, high-effusion outburst eruptions at Rarog and Heno Paterae on Io. Icarus 242, 365-378 (2014).}
\item[]{de Pater, I., Davies, A.G., Marchis, F. Keck observations of eruptions on Io in 2003-2005. Icarus 274, 284-296 (2016).}
\item[]{Gaskell, R.W., Synnott, S.P., McEwen, A.S., Schaber, G.G. Large-scale topography of Io - Implications for internal structure and heat transfer. GRL 15, 581-584 (1988).}
item[]{Hodapp, K.W., Jensen, J.B., Irwin, E.M., et al. The Gemini near-infrared imager (NIRI). PASP 115, 1388-1406 (2003).}
\item[]{Koga, R., Tsuchiya, F., Kagitani, M., et al. The time variation of atomic oxygen emission around Io during a volcanic event observed with Hisaki/EXCEED. \textit{Icarus} 299, 300-307 (2018).}
\item[]{Lopes-Gautier, R., McEwen, A.S., Smythe, W.B., et al. Active volcanism on Io: Global distribution and variations in activity. \textit{Icarus} 140, 243-264 (1999).}
\item[]{Hamilton, C.W. et al. Spatial distribution of volcanoes on Io: Implications for tidal heating and magma ascent. Earth Planet Sci Lett 361, 272-286 (2013).}
\item[]{Ivezi\'c, \v{Z}., Connolly, A.J., VanderPlas, J.T., Gray, A. \textit{Statistics, Data Mining, and Machine Learning in Astronomy}, Princeton Series in Modern Observational Astronomy, pp. 140-144 (2014).}
\item[]{Johnson, T.V., Veeder, G.J., Matson, D.L. et al. Io: Evidence for silicate volcanism in 1986. \textit{Science} 242, 1280-1283 (1988).}
\item[]{Marchis, F., de Pater, I., Davies, A.G. et al. High-resolution Keck adaptive optics imaging of violent volcanic activity on Io. Icarus 160, 124-131 (2002).}
\item[]{Masters, D.C., Stern, D.K., Cohen, J.G. et al. The Complete Calibration of the Color-Redshift Relation (C3R2) Survey: Survey Overview and Data Release 1. ApJ, 841, 111, 10pp (2017).}
\item[]{Morgenthaler, J.P., Rathbun, J.A., Schmidt, C.A., Baumgardner, J., Schneider, N.M. Large volcanic event on Io inferred from jovian sodium nebula brightening. ApJ Lett 871, article id. L 23, 6pp (2019).}
\item[]{Rathbun, J.A., Spencer, J.R. Ground-based observations of time variability of multiple active volcanoes on Io. Icarus 209, 625-630 (2010).}
\item[]{Rathbun, J.A., Howell, R.R., Spencer, J.R. Active volcanoes on Io: Putting ground-based observations of Jupiter occultations into the PDS. 49th LPSC \#2083 (2018).}
\item[]{Scargle, J.D. Studies in astronomical time series analysis. II - Statistical aspects of spectral analysis of unevenly spaced data. ApJ 263, 835-853 (1982).}
\item[]{Segatz, M. et al. Tidal dissipation, surface heat flow, and figure of viscoelastic models of Io. Icarus 75, 187-206 (1988).}
\item[]{Spencer, J.R., Shure, M.A., Ressler, M.E., et al. Discovery of hotspots on Io using disk-resolved infrared imaging. Nature 348, 618-621 (1990).}
\item[]{Tsang, C.C.C., Rathbun, J.A., Spencer, J.R., Hesman, B.E., Abramov, O. Io's hot spots in the near-infrared detected by LEISA during the New Horizons flyby, \textit{JGR Planets}, 119, 2222-2238 (2014).} 
\item[]{VanderPlas, J.T. Understanding the Lomb-Scargle Periodogram. \textit{ApJ Supp.}, 236:16, 28pp (2018).}
\item[]{Veeder, G.J., Matson, D.L., Johnson, T.V., Blaney, D.L., Goguen, J.D. Io's heat flow from infrared radiometry: 1983-1993. JGR Planets 99 (E8), 17095-17162 (1994).}
\item[]{Veeder, G.J., Davies, A.G., Matson, D.L., Johnson, T.V., Williams, D.A. and Radebaugh, J. Io: Volcanic thermal sources and global heat flow. \textit{Icarus} 219, 701-722 (2012).}
\item[]{Veeder, G.J., et al., 2015. Io: Heat flow from small volcanic features. Icarus 245, 379-410.}
\item[]{Williams, D.A. et al. Geologic map of Io, USGS Scientific Investigations Map 3168, scale 1:15,000,000 (2011a).}
\item[]{Williams, D.A., Keszthelyi, L.P., Crown, D.A. Volcanism on Io: New insights from global geologic mapping. Icarus 214, 91-112 (2011b).}
\item[]{Wizinowich, P., Acton, D.S., Shelton, et al. First light adaptive optics images from the Keck II telescope: A new era in high angular resolution imagery. PASP 112, 315-319 (2000).}
\item[]{Yoshikawa, I., Suzuki, F., Hikida, R., et al. Volcanic activity on Io and its influence on the dynamics of the jovian magnetosphere observed by EXCEED/Hisaki in 2015. \textit{Earth, Planets and Space--Frontier Letter} 69, 110, 11pp (2017).}
\item[]{Zechmeister, M. and K\"urster, M., 2009. The generalized Lomb-Scargle periodogram. A\&A 496, 577-584 (2009).}
\end{itemize}
\clearpage
\setcounter{table}{0}
\renewcommand*\thetable{\Alph{section}.\arabic{table}}
\appendix
\section{Tables}
Table \ref{tbl:obs} provides details on the observations, and Table \ref{tbl:hsphot} provides the measured intensities for all hot spot detections in 2013-2018, corrected for geometric foreshortening. Both Tables \ref{tbl:obs} and \ref{tbl:hsphot} are available for download. 
\startlongtable
\begin{deluxetable}{lllll}
\tabletypesize{\scriptsize}
\tablecaption{Observations\label{tbl:obs}}
\tablehead{\colhead{Date$^a$} & \colhead{Tel/Inst} & \colhead{Lon$^b$} & \colhead{Lat$^b$} & \colhead{Distance$^b$} \\
 \colhead{UT} & \colhead{} & \colhead{\degr W} & \colhead{\degr N} & \colhead{AU}}
\startdata
2013-08-15 & Keck/NIRC2 & 337.0 & 2.0 & 5.85 \\
2013-08-20 & Keck/NIRC2 & 275.0 & 2.0 & 5.79 \\
2013-08-21 & Keck/NIRC2 & 115.0 & 2.0 & 5.78 \\
2013-08-22 & Keck/NIRC2 & 319.0 & 2.0 & 5.77 \\
2013-08-23 & Keck/NIRC2 & 161.0 & 2.0 & 5.76 \\
2013-08-29 & Gemini N/NIRI & 305.7 & 1.9 & 5.69 \\
2013-08-30 & Gemini N/NIRI & 146.2 & 1.9 & 5.68 \\
2013-09-01 & Gemini N/NIRI & 192.7 & 1.9 & 5.65 \\
\enddata
\tablenotetext{a}{Data from 2013-08-15 to 2013-08-23 were previously published in de Pater et al. (2014); and data from 2013-2015 were previously published in de Kleer and de Pater (2016a).}
\tablenotetext{b}{Sub-observer latitude and longitude, and Earth-Io distance, from JPL Horizons.}
\end{deluxetable}
\startlongtable
\begin{deluxetable}{l | l | ll | llllllll}
\tabletypesize{\scriptsize}
\tablecaption{Hot Spot Intensities$^a$\label{tbl:hsphot}}
\tablehead{\colhead{Hot Spot} & \colhead{Date} & \colhead{Lat} & \colhead{Lon} & \colhead{K-cont} & \colhead{H$_2$O} & \colhead{PAH} & \colhead{L'} & \colhead{Br$\alpha$-cont} & \colhead{Br$\alpha$} & \colhead{Ms} \\
\colhead{} & \colhead{} & \colhead{} & \colhead{} & \colhead{$\lambda =$ 2.27 $\mu$m} & \colhead{3.06 $\mu$m} & \colhead{3.29 $\mu$m} & \colhead{3.78 $\mu$m} & \colhead{3.99 $\mu$m} & \colhead{4.05 $\mu$m} & \colhead{4.67 $\mu$m} \\
\colhead{} & \colhead{[UT]} & \colhead{[$^{\circ}$N]} & \colhead{[$^{\circ}$W]} & \colhead{} & \colhead{} & \colhead{} & \colhead{[GW/$\mu$m/sr]} & \colhead{} & \colhead{} & \colhead{} }
\startdata
Nusku Patera & 2017-May-27 & -65.0$\pm$1.1 & 6.2$\pm$1.9 &  &  &  &  &  & 4.0$\pm$1.0 &  &  \\
 & \textbf{Average} & \textbf{-65.0} & \textbf{6.2} & \\
\hline
Uta & 2013-Dec-03 & -33.8$\pm$2.0 & 20.1$\pm$2.1 &  &  &  & 2.7$\pm$1.3 &  &  &  &  \\
 & 2014-Jan-20 & -34.9$\pm$0.8 & 20.9$\pm$0.8 &  &  & 1.6$\pm$0.2 & 3.1$\pm$0.5 &  &  & 5.4$\pm$0.8 &  \\
 & 2014-Feb-10 & -34.1$\pm$0.5 & 23.4$\pm$2.1 &  &  &  & 5.1$\pm$2.6 &  &  & 4.6$\pm$0.9 &  \\
 & 2014-Mar-07 & -34.9$\pm$2.0 & 25.9$\pm$2.2 &  &  &  & 2.2$\pm$1.1 &  &  &  &  \\
 & 2014-Mar-14 & -34.7$\pm$2.0 & 24.8$\pm$2.1 &  &  &  & 3.3$\pm$0.5 &  &  &  &  \\
 & 2014-Oct-30 & -33.7$\pm$0.8 & 21.2$\pm$0.8 &  &  &  & 3.0$\pm$0.4 &  &  & 5.8$\pm$0.9 &  \\
\enddata
\tablenotetext{a}{All flux densities are corrected for geometric foreshortening.}
\end{deluxetable}

\end{document}